\documentclass[revtex4]{emulateapj}
\usepackage {latexsym}
\usepackage{acronym}
\usepackage{graphicx}
\usepackage{subfigure}
\usepackage{xcolor}
\usepackage{float}
\usepackage{array}
\usepackage{amsmath}
\usepackage{multirow}
\usepackage{ulem}
\usepackage[backref,breaklinks,colorlinks,citecolor=blue]{hyperref}

\def\commitID{commitID: b0d49ac21d672a28cf7ba5aa404576a48e4089ca}
\def\commitDATE{Fri Mar 6 14:16:36 2015 +0000}

\begin{document}
\title{Maximizing the detection probability of kilonovae associated with gravitational wave observations}

\author{Man Leong Chan$^1$, Yi-Ming Hu$^{1,2,3}$, Chris Messenger$^1$, Martin Hendry$^1$ \& Ik Siong Heng$^1$}
\affiliation{$^1$ SUPA, School of Physics and Astronomy, University of
  Glasgow, Glasgow G12 8QQ, United Kingdom\\
 $^2$ Max Planck Institute for Gravitational Physics (Albert Einstein Institute), Callinstra{\ss}e 38, D-30167 Hannover, Germany\\
 $^3$ Tsinghua University, Beijing, China}
\email[Yi-Ming Hu: ]{yiming.hu@aei.mpg.de}

\date{\today}
\date{\commitDATE\\\mbox{\small \commitID}}

\begin{abstract}
Estimates of the source sky location for gravitational wave signals are likely
span areas ranging up to hundreds of square degrees or more, making it very
challenging for most telescopes to search for counterpart signals in the
electromagnetic spectrum. To boost the chance of successfully observing such
counterparts, we have developed an algorithm which optimizes the number of
observing fields and their corresponding time allocations by maximizing the
detection probability.  As a proof-of-concept demonstration, we optimize
follow-up observations targeting kilonovae using telescopes including
CTIO-Dark Energy Camera, Subaru-HyperSuprimeCam, Pan-STARRS and Palomar
Transient Factory. We consider three simulated gravitational wave events with
$90\%$ credible error regions spanning areas from ${\sim}30\,\mathrm{deg^2}$ to
${\sim}300\,\mathrm{deg^2}$.  Assuming a source at $200\,\mathrm{Mpc}$, we
demonstrate that to obtain a maximum detection probability, there is an
optimized number of fields for any particular event that a telescope should
observe.  To inform future telescope design studies, we present the maximum
detection probability and corresponding number of observing fields for a
combination of limiting magnitudes and fields-of-view over a range of
parameters. We show that for large gravitational wave error regions, telescope
sensitivity rather than field-of-view, is the dominating factor in maximizing
the detection probability.  
\end{abstract}

\maketitle

\acrodef{GW}[GW]{gravitational wave}
\acrodef{CBC}[CBC]{compact binary coalescence}
\acrodef{EM}[EM]{electromagnetic}
\acrodef{NS}[NS]{neutron stars}
\acrodef{BNS}[BNS]{binary neutron stars}
\acrodef{NSBH}{neutron star-black hole}
\acrodef{SNR}[SNR]{signal-to-noise ratio}
\acrodef{JWST}[JWST]{James Webb Space Telescope}
\acrodef{FOV}[FOV]{field of view}
\acrodefplural{FOV}[FOV's]{fields of view}
\acrodef{F2Y}[F2Y]{The First Two Years of Electromagnetic Follow-Up with Advanced LIGO and Virgo}
\acrodef{GA}[GA]{Greedy Algorithm}
\acrodef{LM}[LM]{Lagrange Multiplier}
\acrodef{PTF}[PTF]{Palomar Transient Factory}
\acrodef{PS}[PS]{Pan-Starrs}
\acrodef{CCD}[CCD]{Charge-coupled device}
\acrodef{HSC}{Subaru-HyperSuprimeCam}
\acrodef{DEC}{CTIO-Dark Energy Camera}
\acrodef{LSST}{Large Synoptic Survey Telescope}

%
\section{Introduction}\label{sec:intro}
%
%
The detection of \acp{GW} from the inspiral and merger of binary black-hole
systems~\citep{abbott2016observation,PhysRevLett.116.241103} has marked the
beginning of the \ac{GW} astronomy era. With the advanced interferometric
detectors, \acp{GW} are expected to be observed from a number of additional
source types including the mergers of \ac{BNS}, and \ac{NSBH} systems. For
these other sources the presence of matter from the \ac{NS} components, in such
a violent interaction with its companion, makes it likely that \ac{EM} emission
will be generated.  The joint detection of a \ac{GW} signal and its \ac{EM}
counterpart is therefore of particular interest.  A coincident \ac{EM}
observation together with the detection of \acp{GW} from these sources would
lead to a deeper and more comprehensive understanding of these objects.  A
successful \ac{EM} follow-up observation triggered by a \ac{GW} event can
greatly improve the identification and localization of the host galaxy, bring
us a more accurate estimate of the distance and energy involved, and provide
better understanding of the merger's hydrodynamics and the local environment of
the progenitor~\citep{2009astro2010s20B,2009aaxoconf312K,2009astro2010S235P}.
Additionally, knowledge obtained by \ac{EM} follow-up observations may break
the modeling degeneracies of binary properties, and confirm the association
between the \ac{GW} source and its \ac{EM} counterpart.  Also, successfully
locating the \ac{EM} counterpart of a \ac{GW} event could increase the
confidence in the \ac{GW} detection~\citep{Blackburn2015}.

%
Among many other potential \ac{EM} counterparts of \ac{GW} events, it has been
argued that kilonovae are a strong candidate for \ac{GW} signals from \ac{BNS}
and \ac{NSBH} mergers~\citep{Li1998,
Metzger2012,Tanvir2013,2016arXiv160100180L}.  We therefore focus our discussion
on kilonovae only.  Kilonovae are predicted to produce an optical/infrared and
isotropic quasi-thermal transient, and thought to originate from the hot
neutron-rich matter ejected from \ac{BNS} mergers. This ejection triggers
r-process nucleosynthesis, the radioactive decay of which sustains
the high temperature of the ejecta and powers the kilonovae.  Kilonovae can be
${\sim}1000$ times brighter than a nova (peak luminosity $L\sim10^{40}\,
\mathrm{erg/s}$)~\citep{Metzger2014}, but given the expected rate of \ac{BNS}
events, their corresponding distances make kilonovae relatively dim \ac{EM}
transients.  In addition, according to~\citet{Metzger2012}, a kilonova
maintains its peak luminosity only for hours to days after merger. However,
more recent calculations including the opacity of the $r-process$
elements~\citep{barnes2013effect,0004-637X-780-1-31,Grossman21032014} indicate
that this timescale is likely to be days to weeks.  Currently, there are three
detections associated with the short-duration gamma-ray bursts GRB
130603B~\citep{Tanvir2013,Berger2013} and GRB
060614~\citep{Yang2015,jin2015light} and GRB 050709~\citep{jin2016050709}. 

%
The sky location estimate for a \ac{GW} event can cover a large fraction of the
sky (${\gtrsim}100\,\mathrm{deg^2}$~\citep{Singer2014}), posing a significant
challenge to telescopes with \acp{FOV} of order ${\sim}1\,\mathrm{deg^2}$ trying
to find counterpart signals in the \ac{EM} spectrum.  Observing and confidently
detecting kilonovae demands long exposure times even for powerful telescopes.
It has been proposed that the infrared sensitivity of the planned \ac{JWST} may
enable detections with short exposure times, thus compensating for its small
\ac{FOV}~\citep{Bartos2015}. The use of galaxy catalogs may also provide prior
information on the direction of an
event~\citep{Fan2014,Bartos2014,Gehrels2015a}.  Nonetheless, even with
\ac{JWST} and galaxy catalogs, the observation time can still span an entire
night, which may make target-of-opportunity \ac{BNS} merger observations less
attractive.

%
Given that current and future \ac{EM} telescopes will be subject to limited
observational resources, we might ask how can we maximize the detection
probability of a \ac{GW} triggered kilonova signature. We have developed an
algorithm to answer this question, and here we present the results of a
proof-of-concept demonstration applied to four different telescopes including
\ac{HSC}\footnote{\href{http://www.subarutelescope.org/Observing/Instruments/HSC/}{http://www.subarutelescope.org/Observing/Instruments/HSC/}}~\citep{miyazaki2012hyper}
,
\ac{DEC}~\citep{bernstein2012supernova},~Pan-Starrs\footnote{\href{http://pan-starrs.ifa.hawaii.edu}{http://pan-starrs.ifa.hawaii.edu}}~\citep{2002SPIE.4836..154K},
and \ac{PTF}~\citep{law2009palomar} for three different simulated \ac{GW}
events. This algorithm takes as inputs \ac{GW} sky localization information,
and returns a guidance strategy for time allocation and telescope pointing for
a given \ac{EM} telescope.  In this paper we restrict the discussion to
kilonovae from \ac{BNS} mergers and leave the discussion of kilonova from
\ac{NSBH} for future study.  This is in part due to the fact that our chosen
\ac{GW} data set from~\citet{Singer2014} contains only \ac{BNS} mergers.
However, we remind the readers that there is no theoretical restriction
preventing us from applying our algorithm using other \ac{EM} counterpart
models and neither is it restricted to using only \ac{GW} sky maps from
\ac{CBC} events. 

%
This paper is organized as follows: the methodology is introduced in
Section~\ref{sec:maths} and we describe the implementation in
Section~\ref{sec:al}. The results obtained with the algorithm and a discussion
of these results are given in Section~\ref{sec:re}. The possible future
directions of this work are provided in Section~\ref{sec:fud}, and our
conclusions are presented in Section~\ref{sec:ccs}. 

%
\section{Methodology}\label{sec:maths}
%
%
As a proof-of-concept of our general EM follow-up method, here we have chosen
to focus on kilonovae counterpart signatures.  We have also adopted some
simplifications which will be relaxed in future studies. We assume that the
available observation time is short compared to the luminosity variation
timescale of a kilonova.  This is validated by the fact that kilonova's
luminosity variation timescale is estimated to be $\sim$days to a week, which
is longer than a reasonable continuous \ac{EM} observation. This approximation
allows us to assume that a kilonova has constant luminosity during the
observation period.  We also only consider the use of R-band luminosity
information, while the method presented can be extended to other regions of the
\ac{EM} spectrum. In reality, identifying a target \ac{EM} counterpart requires
tracking the object's light curves, leading to several observations of the same
point in the sky over several days. However, in this work we deal with the
problem of detection, rather than identification. We consider only single
observations and calculate optimized pointing directions and durations for a
constrained total observation length and constant luminosity. Subsequent
observations for the purpose of identifying variation within a field can be
achieved by repeating our proposed observing strategy at some later time. 

%
In general, \ac{GW} events could be localized to ${\gtrsim}100\,\deg^2$, which
is much larger than a typical \ac{EM} telescope's \ac{FOV}
(${\lesssim}1\,\deg^2$). We therefore do not consider the telescope's rotation
around its own axis, hence throughout this work, reference to single telescope
pointing implies a rectangular \ac{CCD} image with edges parallel to lines of
longitude and circles of latitude.

%
We also assume that the prior information from a \ac{GW} trigger can be
approximated as having independent sky location and distance probability
distributions. Generally, this may not be the case, however, our mathematical
treatment can be greatly simplified under this assumption. We note that the sky
maps available for our chosen dataset (discussed in Sec.~\ref{sec:al})
naturally lend themselves to this approximation since no distance information
was computed as part of the rapid sky localization study reported
in~\citet{Singer2012}\footnote{The rapid sky localization algorithm BAYESTAR
has since been updated to also provide rapid distance
posteriors~\citep{SingerLP:2016ur}.}. The final assumption is that the \ac{EM}
telescope can see every direction regardless of the location of the Sun, Moon,
and the horizon. The issue of optimization of \ac{EM} follow-up observations
under the time-critical constraints such as those imposed by a source dipping
below the horizon is explored in~\citet{Rana:2016ve}.

%
\subsection{Bayesian Framework}
%
%
We use $D_{\text{EM}}$ to denote the successful detection of an \ac{EM}
counterpart. The probability of this occurring depends on the size of the
selected telescope's \ac{FOV} $\omega$, the observed sky locations$(\alpha,
\delta)$, and the exposure time $\tau$.  The posterior probability of
successful detection is then given by
\begin{equation}\label{eq:prob}
P(D_{\mathrm{EM}}|\omega, \tau,I) = \int_{N^*}^\infty
dN~p(N|\omega,\alpha, \delta,\tau,I).
\end{equation}
Here, $I$ is prior information that includes the selected telescope's
parameters, such as its photon collecting area $\mathrm{A}$, filter and
\ac{CCD} efficiency. For a particular observation the number of photons $N$
collected by the telescope is
\begin{equation}\label{eq:td}
N = 10^{\frac{25-m}{2.5}}\times \left(\tau A\right)
\end{equation}
where $m$ is the apparent magnitude of the observed source. The threshold count
$N^*$ is the criterion for detection determined by the \ac{SNR} threshold,
background noise and the selected telescope's sensitivity.  The value of $N^*$
is given by Eq.~\ref{eq:td} with input values of $m$ and $\tau$ corresponding
to the selected telescope's detection threshold (see
Table~\ref{tab:telescope_sens}).  The constant $10^{10}$ in Eq.~\ref{eq:td} is
the number of photons per second at $m=0$. In practice, the value of $N^*$
should also account for the change in background light accumulated for
different choices of observation time $\tau$, however, for simplicity we ignore
this effect in this work. Since the number of photons expected from a target
\ac{EM} counterpart depends on its absolute magnitude $M$, distance $R$ from
the telescope, and how likely that the \ac{GW} event is located within the
field being observed, Eq.~\ref{eq:prob} can be expanded such that
\begin{eqnarray}\label{eq:ep1}
P(D_{\mathrm{EM}}|\omega,\tau,I) =
\int_{N^*}^\infty dN \int dM \int dR \times \nonumber \\
\int_{\omega} d\alpha\,d\delta~ p(N|M,R,\tau,I)
p(M|I)p(\alpha,\delta,R|I).  
\end{eqnarray}
The quantity $P(N|M,R,\tau,I)$ is the probability of receiving $N$ photons from
a source, given its absolute magnitude $M$, distance $R$, and observation time
$\tau$, and is described by a Poisson distribution. Since we assume that the prior
distribution on the distance to the target \ac{EM} counterpart is statistically
independent of the prior distribution on its sky location, Eq.~\ref{eq:ep1} can
be written as
\begin{equation}\label{eq:ep2}
P(D_{\mathrm{EM}}|\omega,\tau,I)
= P_{\mathrm{GW}}(\omega)\times P_{\mathrm{EM}}(\tau)
\end{equation}
where,
\begin{subequations}
\begin{eqnarray}
P_{\mathrm{GW}}(\omega) &=& \int_{\omega}p(\alpha,\delta|I)d\alpha\,d\delta\label{eq:Pgw},\\
P_{\mathrm{EM}}(\tau) &=& \int dM\int dR\int_{N^*}^\infty dN\nonumber\\
&&p(N|M,R,\tau,I)p(R|I)p(M|I).\label{eq:ep3}
\end{eqnarray}
\end{subequations}
%
%
\begin{table}[]
\centering
\caption{Telescope parameters\label{tab:telescope_sens}}
\label{table:tsen}
\begin{tabular}{lccccc}
\hline\hline
\multirow{3}{*}{Telescope} & Aperture & \ac{FOV} & Exposure & Sensitivity
&$N^*/A$\\
                & (m) & (deg$^{2}$) & (s) &(5-$\sigma$ mag &  \\
                & & & &   in R-band)& (m$^{-2}$)\\
\hline
DEC        & 4.0  & 3.0  & 50           & 23.7                               &
162.0                                     \\
HSC        & 8.2  & 1.13\footnote{The full \ac{HSC} \ac{FOV} is 1.77\,deg$^{2}$
but
$\sim 20\%$ is used for calibration purposes.}   & 30           & 24.5
& 46.5                                      \\
Pan-Starrs & 1.8  & 7.0  & 60           & 22.0                               &
930.5                                     \\
PTF        & 1.2  & 7.0  & 60           & 20.6                               &
3378.3                                    \\
LSST       & 6.7  & 9.6  & 15           & 24.5                               &
23.26                                    \\
\hline\hline
\end{tabular}
\end{table}
%
%
It should be noted that the \ac{GW} sky localization information used for this
work has been marginalized over distance, meaning that the \ac{GW} information
represents a 2D error region projected onto the sky.  However, since the
distance estimate inferred from a \ac{GW} observation will typically be poorly
constrained and can be well approximated by a Gaussian
distribution~\citep{Veitch2014,SingerLP:2016ur}, we assume a Gaussian prior
with mean $= 200\mathrm{Mpc}$ and standard deviation $= 60\mathrm{Mpc}$ for the
distance. In principle any form of positional information can be incorporated
into our analysis and therefore our method can be adapted to include more
realistic \ac{GW} distance information. It is also possible that further
constraints from galaxy catalogs can also be incorporated into our
method~\citep{Fan2014}.

%
We assume the least informative prior on peak luminosity such that $p(L|I)
\propto L^{-\frac{1}{2}}$.  It then follows that the prior on peak magnitude is
given by
\begin{equation}\label{eq:magnitude}
p(M|I)\propto 10^{-\frac{M}{5}}
\end{equation} 
where we assume $M$ has a prior range of $(-13,-8)$ as defined by the peak
magnitudes of the models in~\citet{barnes2013effect}.

%
The probability of \ac{EM} detection as defined in Eq.~\ref{eq:prob} considers
only one observing field.  Given the size of a \ac{GW} sky localization error
region, and the typical size of an \ac{EM} telescope's \ac{FOV}, the number of
fields needed to be considered is ${>}1$. If an error region enclosing $90\%$
of the \ac{GW} probability covers $S\,\mathrm{deg^2}$ and the \ac{EM} telescope
\ac{FOV} is $w\,\mathrm{deg^2}$, the maximum number $n$ of fields\footnote{In
practice, \textit{n} can be slightly larger due to overlapping fields.}
required to cover the error region at $90\%$ can be estimated as
$n{\lesssim}S/\omega$.

%
One might assume that observing as many fields as possible is optimal but we
will show that telescope time is better spent by observing $k$ fields where $k$
lies in the range $[1,n]$. This occurs when it becomes more beneficial to
observe a particular field for longer than observing a new field.

%
For a given total observation time $T$, we are free to choose which fields
we observe and the observation time allocated to each field. We represent these
quantities by the vectors $\{\omega^{(k)}\}$ and $\{\tau^{(k)}\}$ respectively
where $k$ is the total number of chosen fields. Maximizing the detection
probability of a kilonova amounts to finding the values of these vectors and
the value of $k$ which maximizes:
\begin{eqnarray}\label{eq:sum}
P(D_{\mathrm{EM}}|k) &\equiv& P(D_{\mathrm{EM}}|\{\omega^{(k)}\},\{\tau^{(k)}\},I)
\nonumber\\
&\equiv& \sum_{i=1}^{k\leq n} P(D_{\mathrm{EM}}|\omega^{(k)}_{i},\tau^{(k)}_{i},I).
\end{eqnarray}
The choices of $\{\tau^{(k)}\}$ are subject to the constraint that 
\begin{equation}\label{eq:tauconstraints}
kT_{0} + \sum_{i=1}^{k} \tau^{(k)}_{i} = T,
\end{equation}
where $T_0$ represents the time required to slew between telescope pointings
and/or perform \ac{CCD} readout, and is equal to
$\mathrm{max}(\mathrm{slew~time, CCD~readout~time})$. We treat $T_{0}$ as
independent of the angular distance between pointings.

%
The expression we have for kilonova detection probability (Eq.~\ref{eq:sum}) as
a function of the number of observed fields depends on our choice of field
location and observation time within each field. Given a number of fields $k$
we begin choosing fields with a greedy algorithm, which will be described in
section \ref{sec:al}.  Once the $k$ fields have been chosen, they are
represented by $\{\omega^{(k)}\}$ and Eq.~\ref{eq:sum} is maximized over the
parameter vector $\{\tau^{(k)}\}$ to obtain the optimal kilonova detection
probability. This is then repeated for each $k$ in the range $[1,n]$ to find
the optimal number of observed fields $k$.

%
%
\section{Implementation}\label{sec:al}
\begin{figure}
\centering
  \includegraphics[width=1\columnwidth]{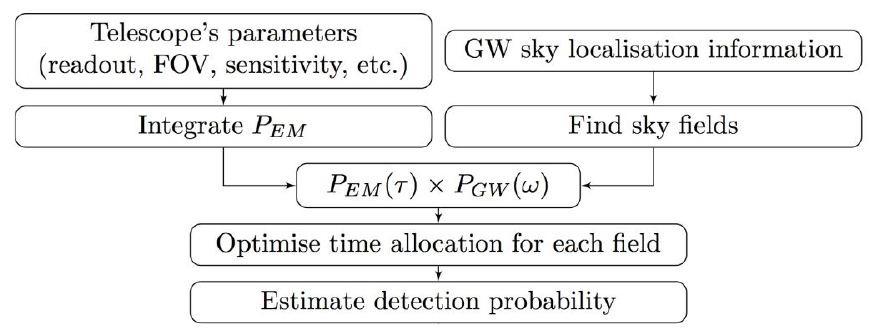}
\caption{The process for generating an optimized observing strategy. Our
algorithm takes two inputs: the \ac{GW} sky localization information and
a set of telescope parameters. After integrating over the number of received
photons $N$, the source distance $R$, and the source absolute magnitude $M$, the
algorithm returns the probability of detection of an \ac{EM} kilonova signal
$P_{\mathrm{EM}}$ as a function of the field observation time $\tau$. In
parallel, for each choice of the total number of observed fields $k$, the
algorithm selects the fields using a greedy algorithm. Based on the enclosed
\ac{GW} probability within each field the corresponding optimized observation times
are computed using a Lagrange Multiplier approach. The total \ac{EM} detection
probability is then output for each choice of {k}.
\label{fig:fc}}
\end{figure}

%
%
In this section we describe the processes for applying the \ac{GW} sky
localization information and generating the optimal observing strategy. The
flow chart in Fig.~\ref{fig:fc} is a visual representation of this process.

%
As shown in Fig.~\ref{fig:fc}, we require information regarding the sky
position of the \ac{GW} source which we obtain from the BAYESTAR
algorithm~\citep{Singer2015:Bayestar}. This algorithm outputs \ac{GW} sky
localization information using a
HEALPix\footnote{\href{http://healpix.sourceforge.net}{http://healpix.sourceforge.net}}
coverage of the sky. Each HEALPix point corresponds to a value of the \ac{GW}
probability and represents an equal area of the sky.  BAYESTAR can rapidly (in
${\sim}10$s) generate event location information and has been shown
in~\citet{Singer2014} to closely match results from more computationally
intensive off-line Bayesian inference methods~\citep{Veitch2014}. The simulated
\ac{GW} events used in this work are from \ac{BNS} systems and taken directly
from the dataset used in~\citet{Singer2014}.

%
In this work, we consider follow-up observations using four telescopes
(\ac{HSC},~\ac{DEC},~Pan-Starrs, and \ac{PTF}) for three simulated representative
\ac{GW} events (see Table~\ref{table:1}) which are studied assuming three total
observation times $T=2,4,6\,\mathrm{hrs}$.  For any telescope, \ac{GW} event
and total observation time the first stage of our procedure is to calculate the
maximum number of fields $n$ required to cover the sky area enclosed by the
$90\%$ probability contour of the \ac{GW} sky region.

%
In order to identify the possible observing fields for a given \ac{GW} event,
the greedy algorithm identifies the least number of HEALPix locations on the
sky whose sum of \ac{GW} probability is equal to the desired confidence level
(in our case $90\%$).  Then, assuming that each of those points represents the
center of an observing field we compute the sum of \ac{GW} probability from
each HEALPix point whose center lies within each of those fields.  The field
returning the maximum sum of the \ac{GW} probability among those fields will be
the first field.  Subsequent fields are found by the same procedure, with the
HEALPix points in the previous fields ignored. The summed probability within
each field is an accurate approximation to the quantity $P_{\text{GW}}(w)$ as
defined in Eq.~\ref{eq:Pgw}. The $n$ selected fields are labeled in the order
with which they are chosen and hence their label indicates their rank in terms
of enclosed \ac{GW} probability. Therefore the first $k\leq n$ fields represent
our optimized choice of the values of $\{\omega^{(k)}\}$ in Eq.~\ref{eq:sum}.

%
As shown in Eqns.~\ref{eq:ep2},~\ref{eq:ep3} and~\ref{eq:sum}, the detection
probability achieved by observing $k$ selected fields can be expressed as the
sum of the product of the \ac{EM} and \ac{GW} probabilities in each field. For
each value of $k$ in the range $[1,n]$ we apply a Lagrange Multiplier to find
the solution for the values $\{\tau^{(k)}\}$ that maximizes the detection
probability given by Eq.~\ref{eq:sum}. This is subject to the constraints
defined in Eq.~\ref{eq:tauconstraints} and the value of $k$ that returns the
highest detection probability is identified as the optimal solution. The
analysis therefore guides us as to which subset of fields should be observed
with the selected telescope and how much time should be allocated to each of
those selected fields given the total observation time constraint.

%
\section{Results}\label{sec:re}
%
%
In this section we present the results of our algorithm using our four example
telescopes applied to the follow up of three simulated \ac{GW} events. These
events are taken from the dataset used in~\citep{Singer2014} and are designated
with the IDs 28700, 19296 and 18694. The error regions for these events each
cover ${\sim}300\,\mathrm{deg^2}$, ${\sim}100\,\mathrm{deg^2}$, and
${\sim}30\,\mathrm{deg^2}$ respectively and details of these events are presented
in Table~\ref{table:1}. These events were chosen to represent the potential
variation in sky localization ability of a global advanced detector network. We
highlight that the actual injected distances for our chosen events are
$51\,\mathrm{Mpc}$, $27\,\mathrm{Mpc}$ and $12\,\mathrm{Mpc}$ respectively, while for
our analysis we have assumed a distance of $200\,\mathrm{Mpc}$ for each event.
However, the validity of this work is not undermined. The injected events were
originally simulated assuming a 2 detector aLIGO First Observational Run
configuration. Our analysis assumes a 2(3) detector design sensitivity aLIGO
configuration. The \ac{SNR} and sky localization for an event at
$200\,\mathrm{Mpc}$ in the latter configuration is comparable (to within
factors of ${\sim}$few) to events at a few
tens $\mathrm{Mpc}$ for the former configuration.
\begin{table}
\caption{Simulated \ac{GW} event parameters.\label{table:1}}
\centering
\begin{tabular}{cccccc} 
 \hline\hline
 \multirow{2}{*}{Event ID$^{1}$} & \multirow{2}{*}{SNR} & 90\% Region & Chirp Mass\\ 
   & & $(\mathrm{deg^{2}})$ &  $(M_{\odot})$ \\ 
  \hline
    28700 & 16.8 & 302  &  1.33\\ 
    19296 & 24.3 & 103  &  1.28\\
    18694 & 24.0 & 28.2 &  1.31\\
 \hline
\end{tabular}
\footnotetext{$^{1}$ The Event ID is that given to the events used
in~\citep{Singer2014}}
\end{table}

\begin{figure*}
  \centering
    \subfigure[$90\%$ coverage tilings used for the \ac{HSC} telescope.]
    {
      \includegraphics[width=.33\textwidth]{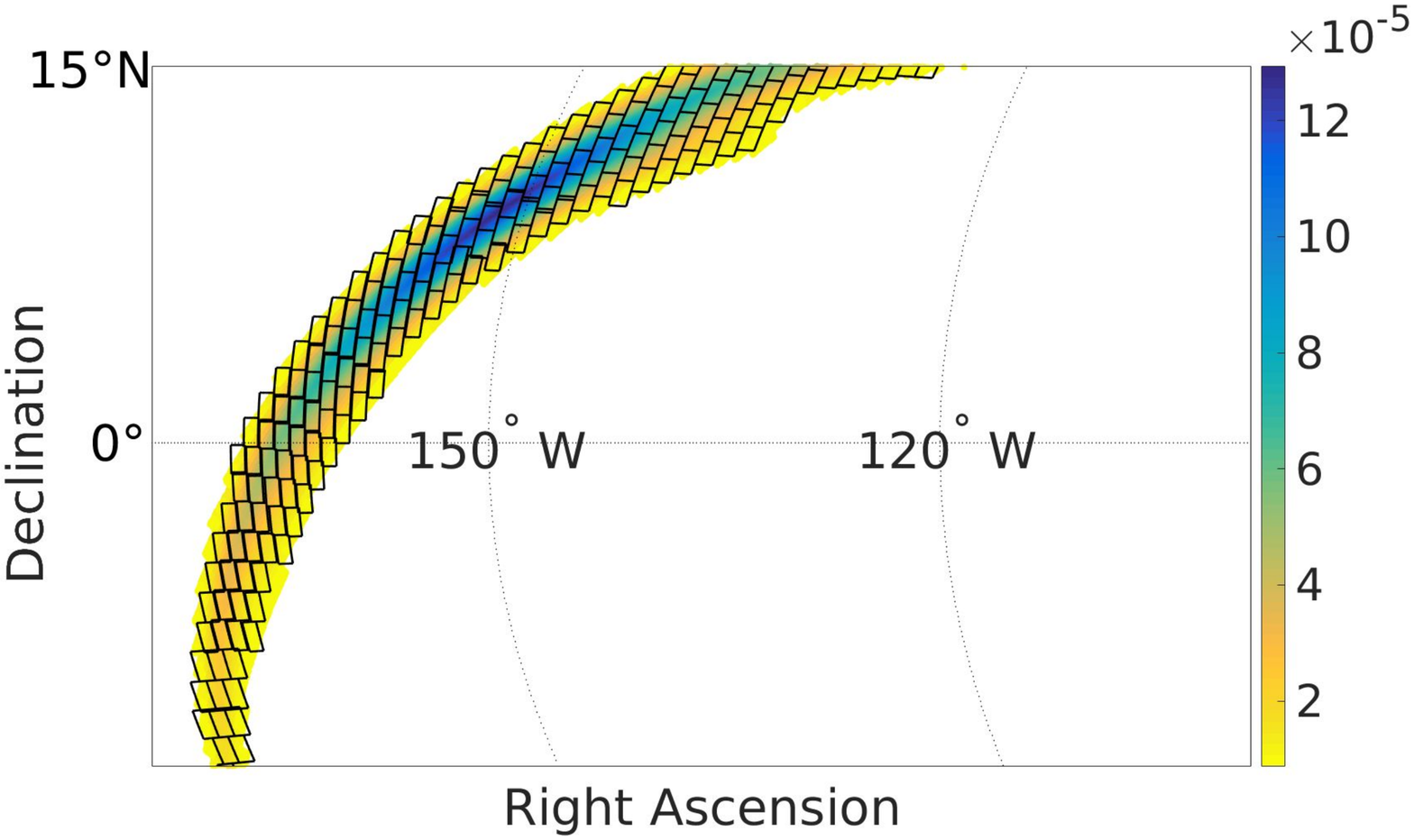}
      \includegraphics[width=.33\textwidth]{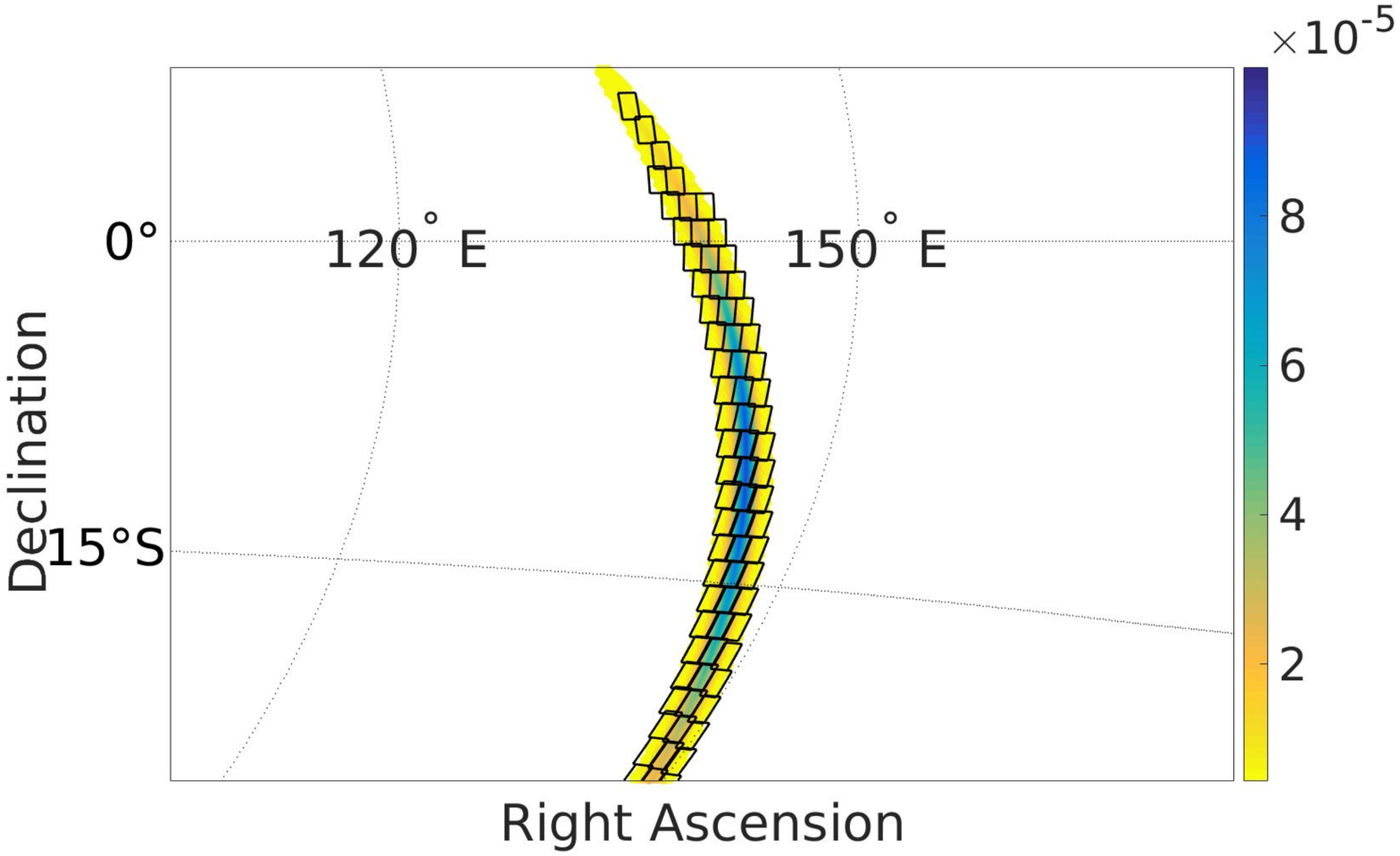}
      \includegraphics[width=.33\textwidth]{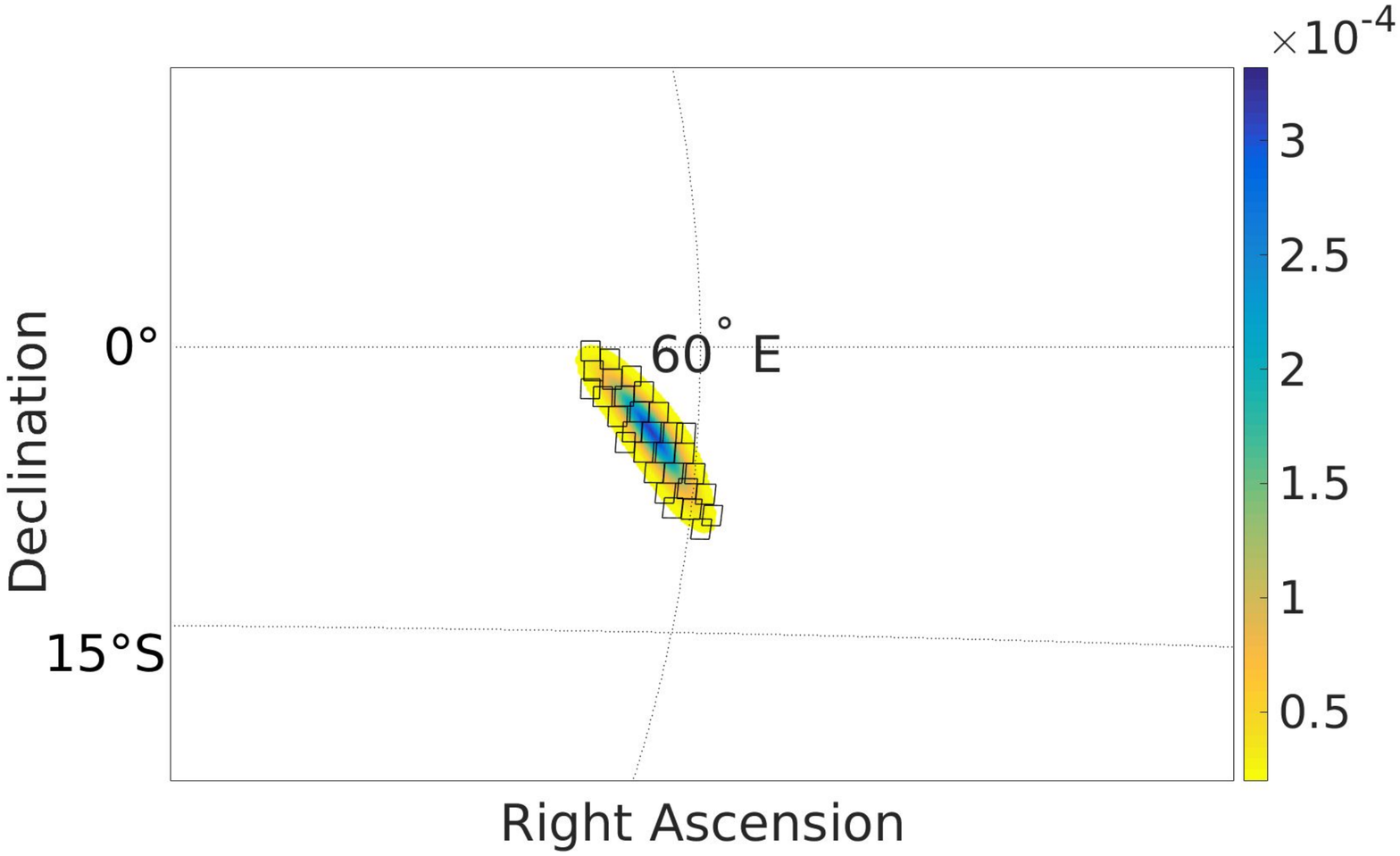}
    }
    \subfigure[$90\%$ coverage tilings used for the \ac{DEC} telescope.]
    {
      \includegraphics[width=.33\textwidth]{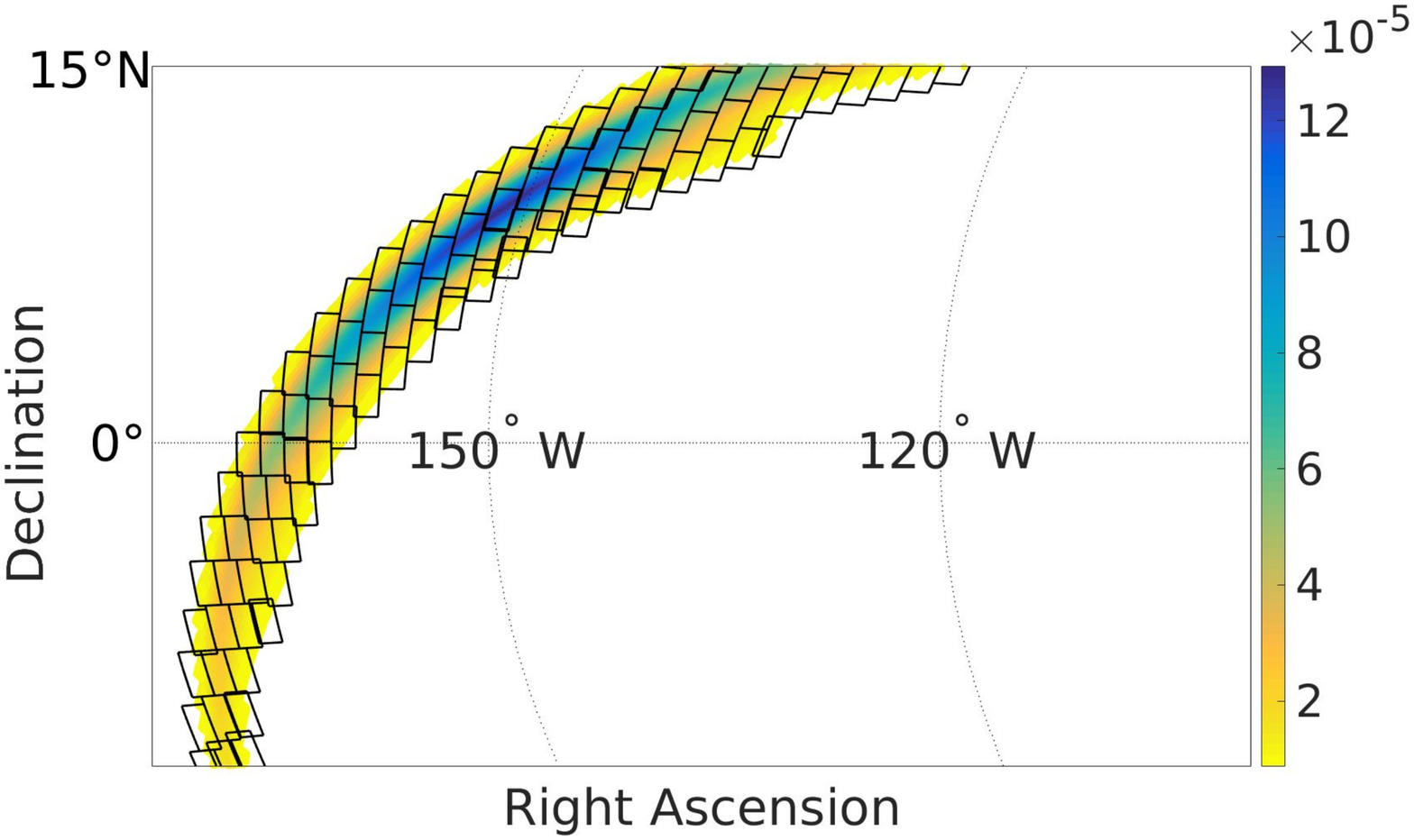}
      \includegraphics[width=.33\textwidth]{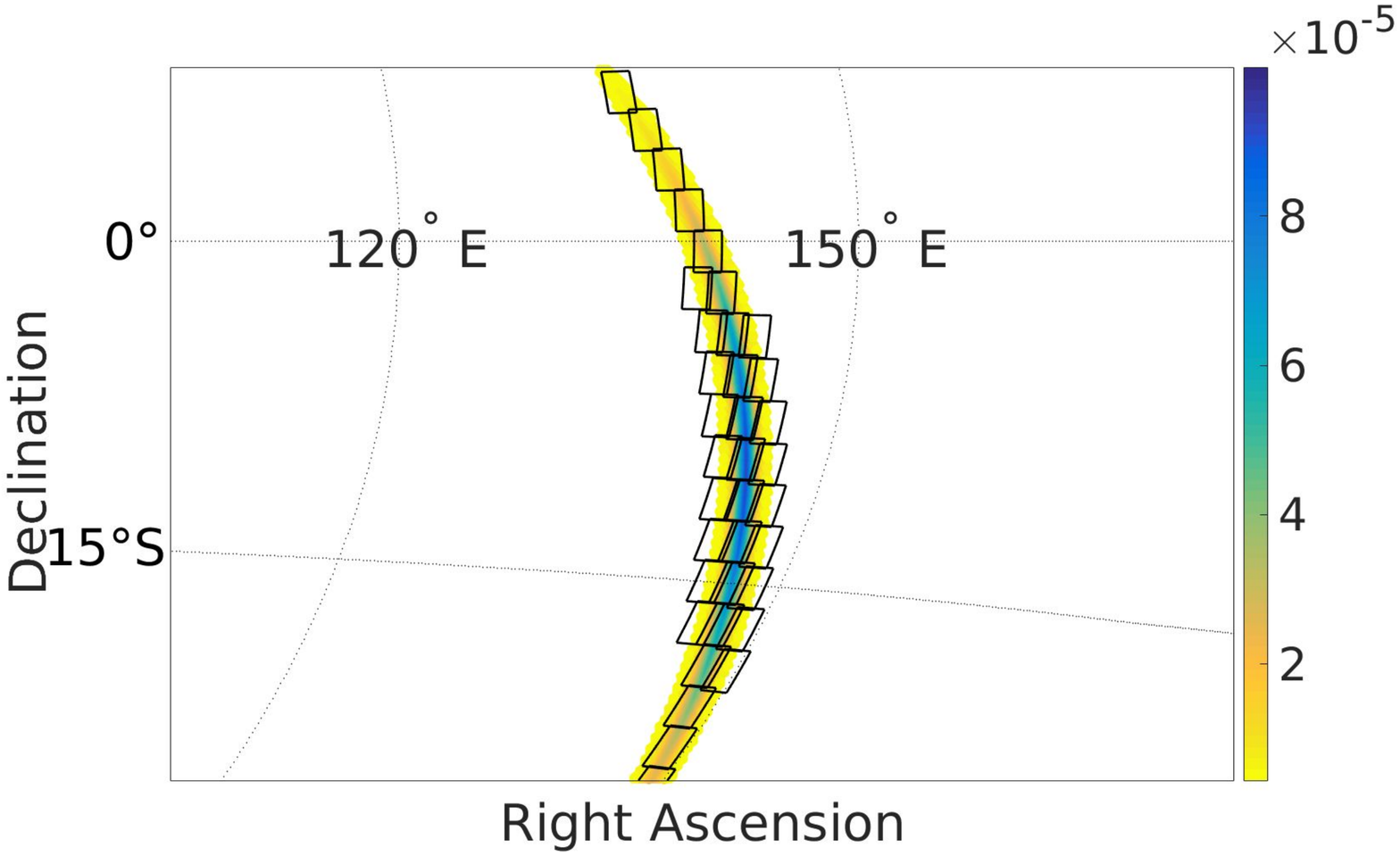}
      \includegraphics[width=.33\textwidth]{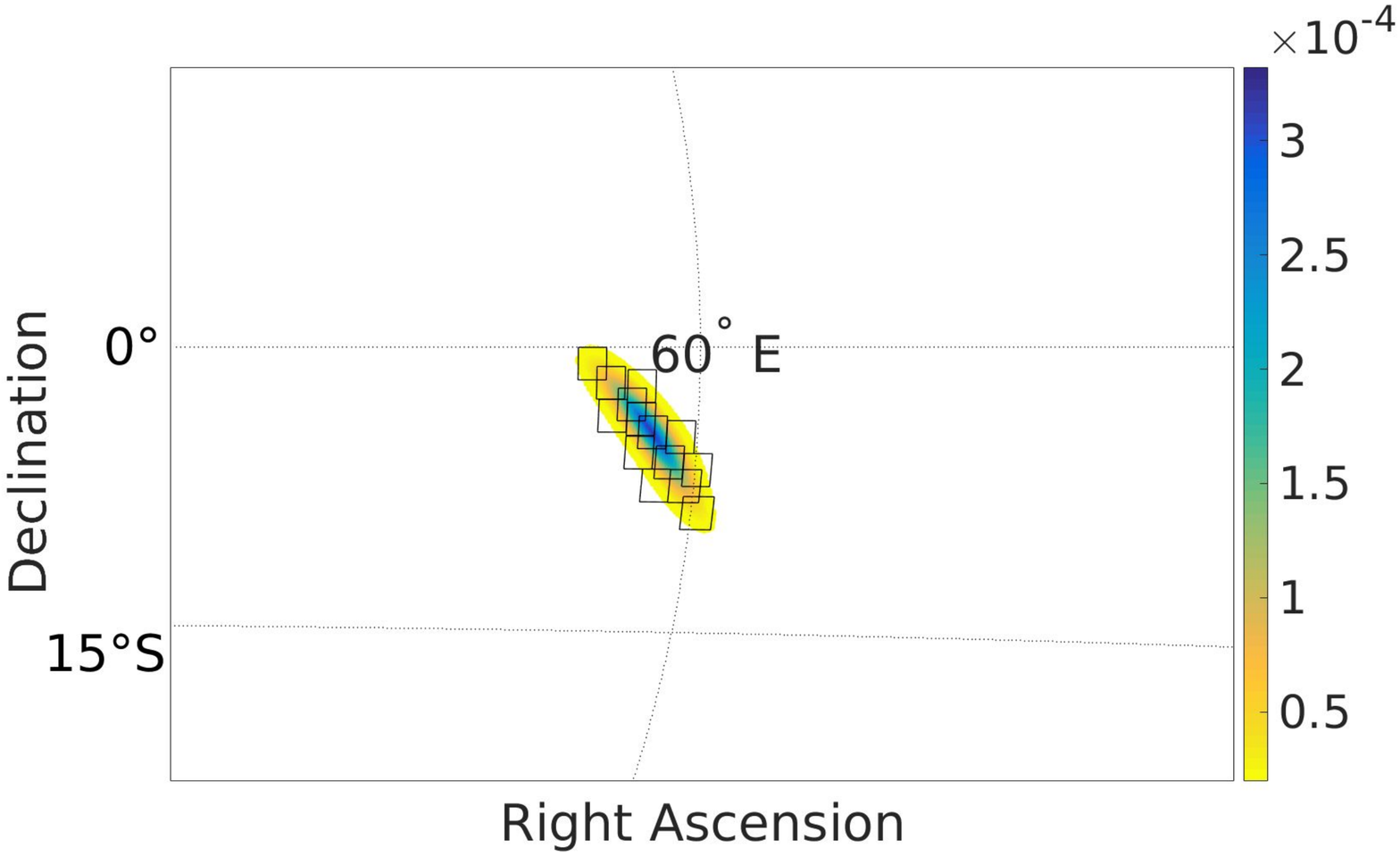}
    }
    \subfigure[$90\%$ coverage tilings used for the Pan-Starrs telescope.]
    {
      \includegraphics[width=.33\textwidth]{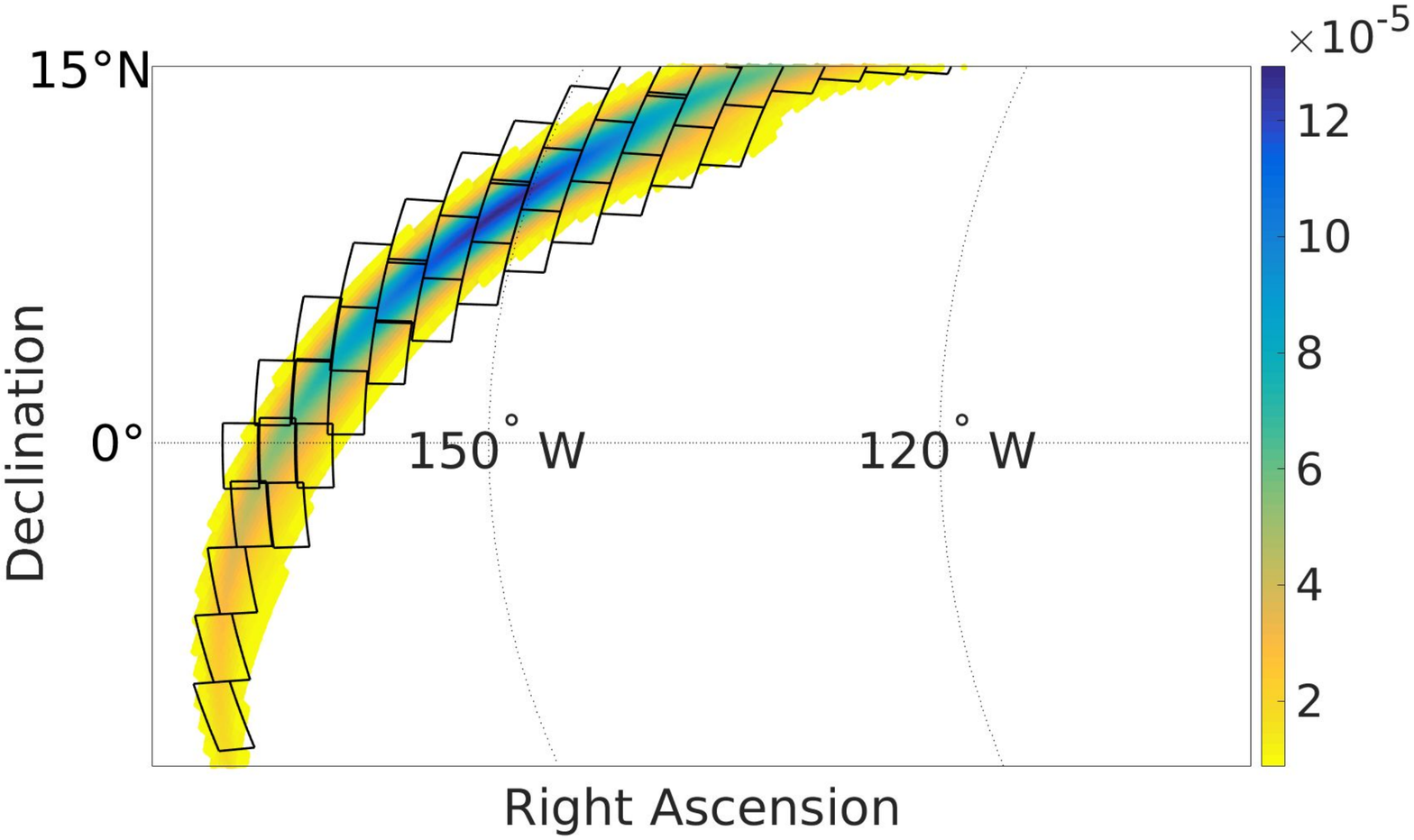}
      \includegraphics[width=.33\textwidth]{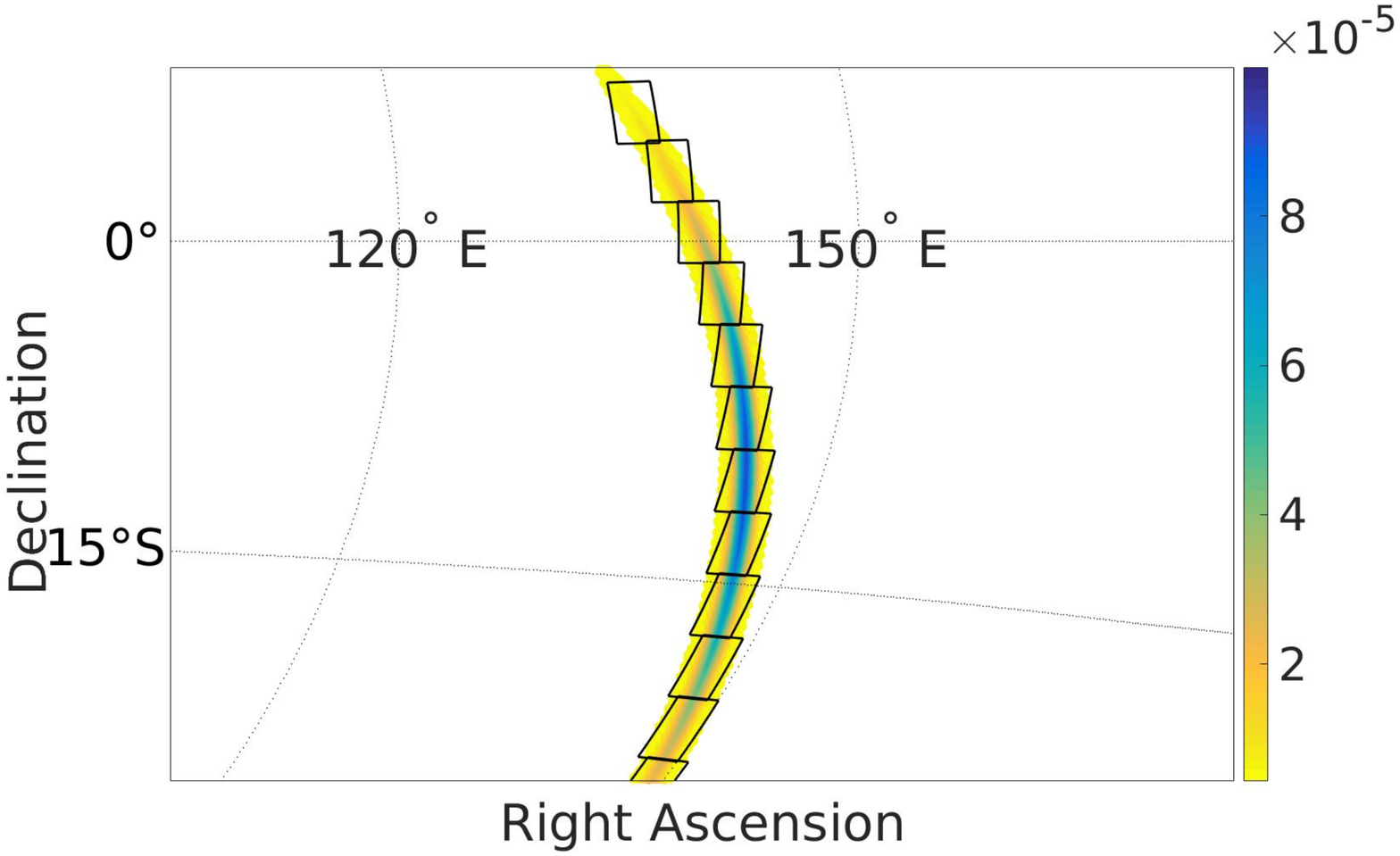}
      \includegraphics[width=.33\textwidth]{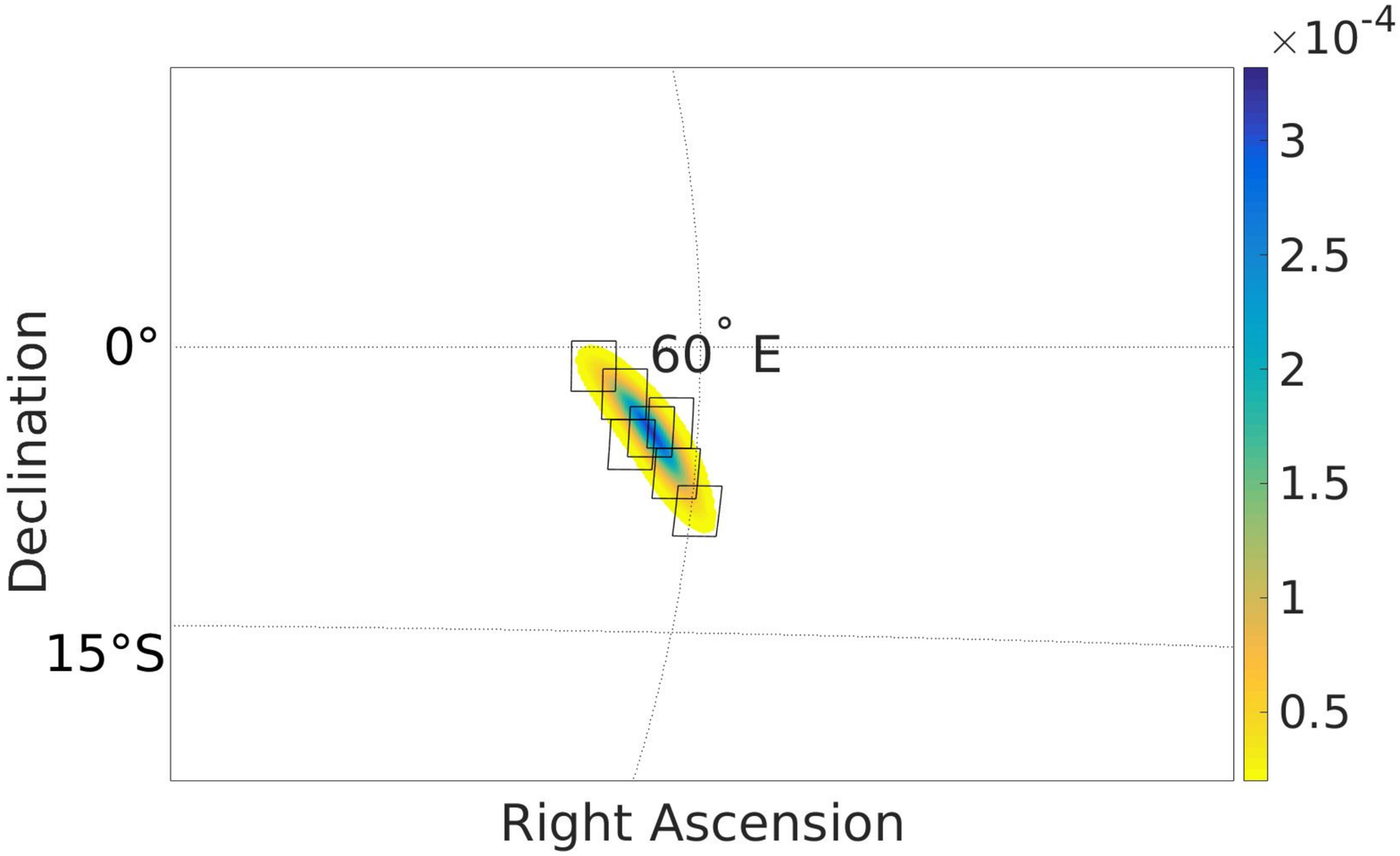}
    }
    \subfigure[$90\%$ coverage tilings used for the \ac{PTF} telescope.]
    {
      \includegraphics[width=.33\textwidth]{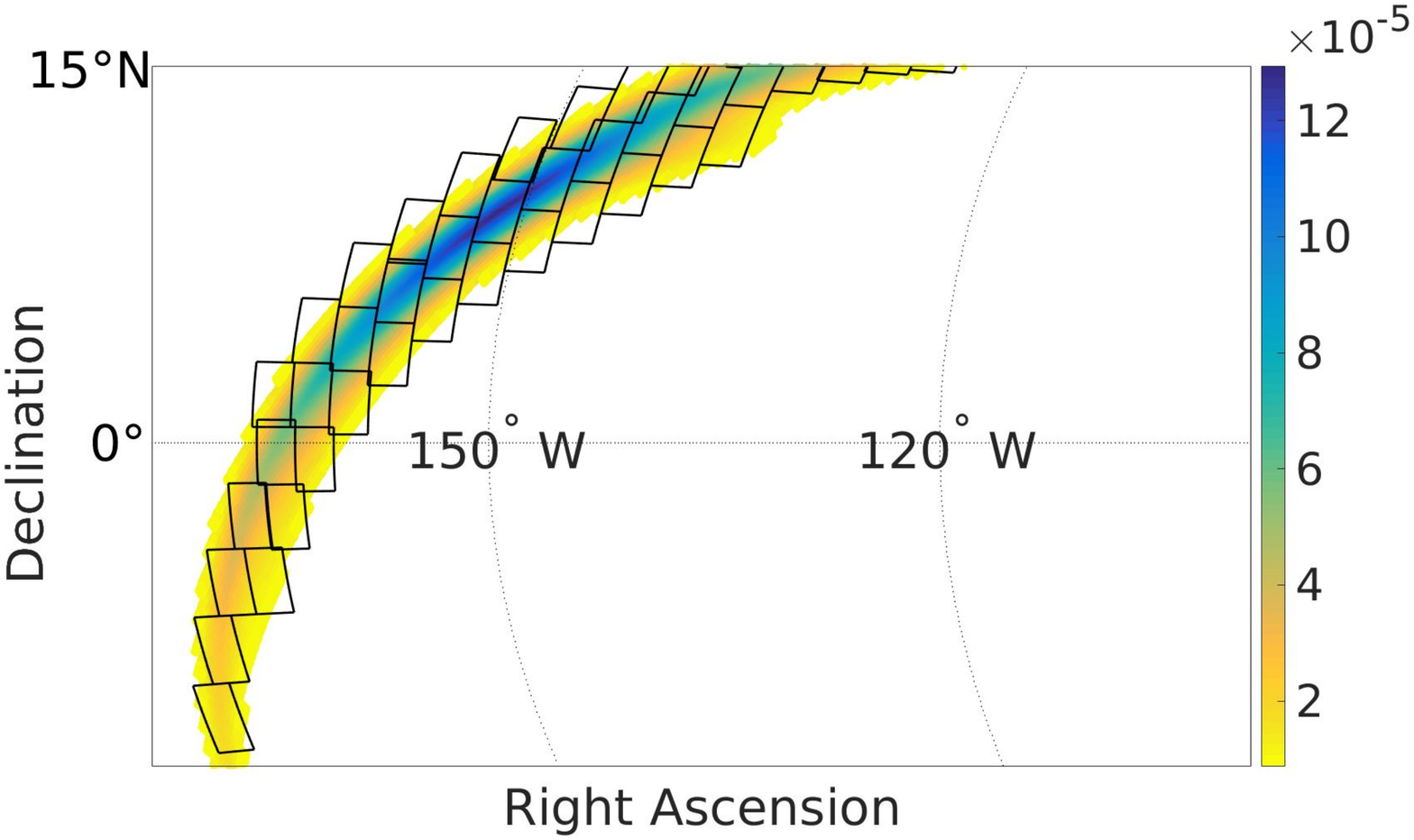}
      \includegraphics[width=.33\textwidth]{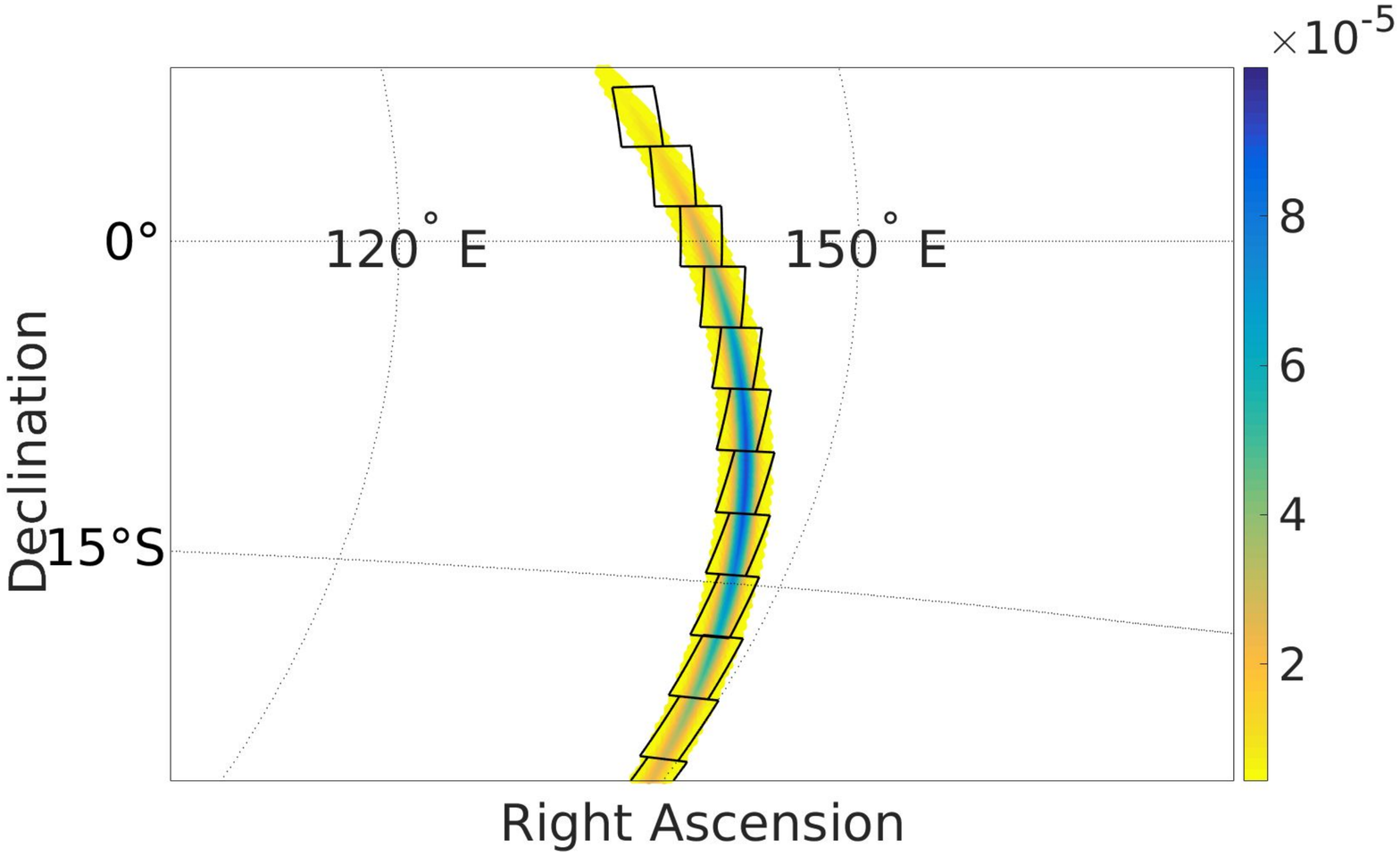}
      \includegraphics[width=.33\textwidth]{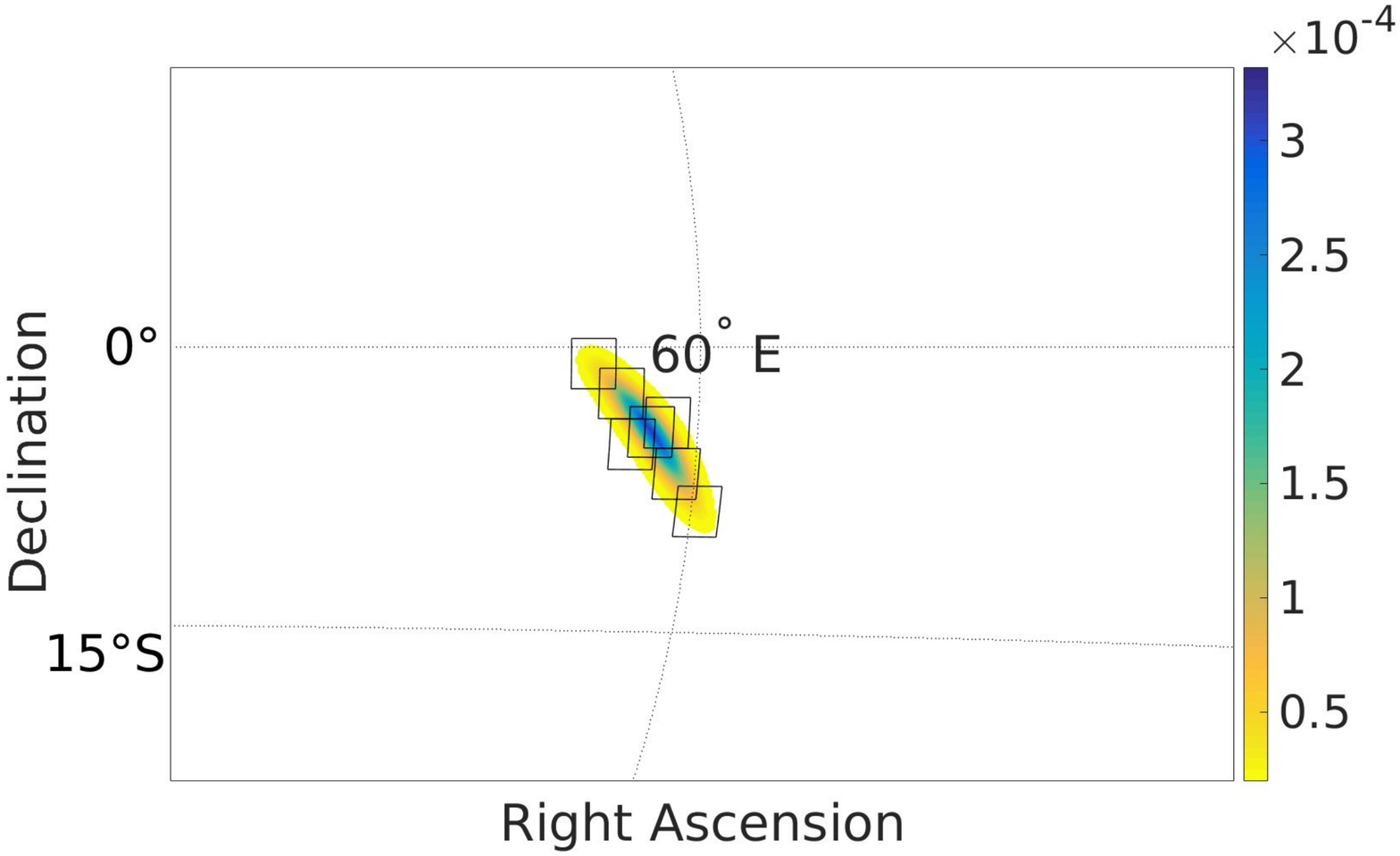}
    }
    \caption{The optimized locations of the observing fields covering $90\%$ of the
\ac{GW} probability for each telescope and for each of the 3 simulated \ac{GW}
events. Each row of plots corresponds to the 4 different telescopes
(a)~\ac{HSC}, (b)~\ac{DEC}, (c)~Pan-Starrs, and (d)~\ac{PTF}.  Within each row
we show the optimized observing field locations for the
${\sim}300\,\text{deg}^{2}$ event (ID 28700) on the left, the
${\sim}100\,\text{deg}^{2}$ event (ID 19296) in the center, and the
${\sim}30\,\text{deg}^{2}$ event (ID 18694) on the right. In each plot the
\ac{GW} sky error is shown as a shaded region with the color bar indicating the
value of posterior probability density.\label{fig:bulk}}
\end{figure*}

%
Figure~\ref{fig:bulk} shows the optimized tiling of observing fields obtained
using the greedy algorithm approach. For each telescope we show sky maps of the
\ac{GW} probability overlaid with the $90\%$ coverage tiling choices for the
three representative \ac{GW} events. The \acp{FOV} of the telescopes range from
$1.13\,\text{deg}^{2}$ to $7.1\,\text{deg}^{2}$ and as such the required number
and location of tilings differ accordingly.  The largest and smallest number of
observation tilings are 230 and 7 for the largest \ac{GW} error region (ID
28700) using \ac{HSC} and for the smallest error region (ID 18694) using either
Pan-Starrs or \ac{PTF} respectively.    

%
Each of the event maps are the result of an analysis assuming only 2 \ac{GW}
detectors. Without a third detector the sky location of an event is restricted
to a thin band of locations consistent with the single time delay measurement
between detectors. This degeneracy is partially broken with the inclusion of
antenna response information resulting in extended arc structures. The third
event that we consider (ID 18694) has sufficient \ac{SNR} and suitable
orientation with respect to the detector network that even with 2 detectors the
sky region is well localized and is only partially extended. We consider this
event to be approximately representative of sky maps obtained from a 3 detector
network. We also note that given the imperfect duty factors of both the initial
and advanced detectors it is highly likely that future detections will be made
whilst one or more detectors are offline. We therefore use the first 2 example
events (ID 28700 and 19296) as simultaneously representative of such a
2-detector scenario and of the potential situation in which a third detector is
significantly less sensitive than the other two. 

\begin{figure*}
  \centering
    \subfigure[\ac{EM} detection probability and optimized observation
times for $T=6$ hour total observation times.]
    {
      \includegraphics[width=.45\textwidth]{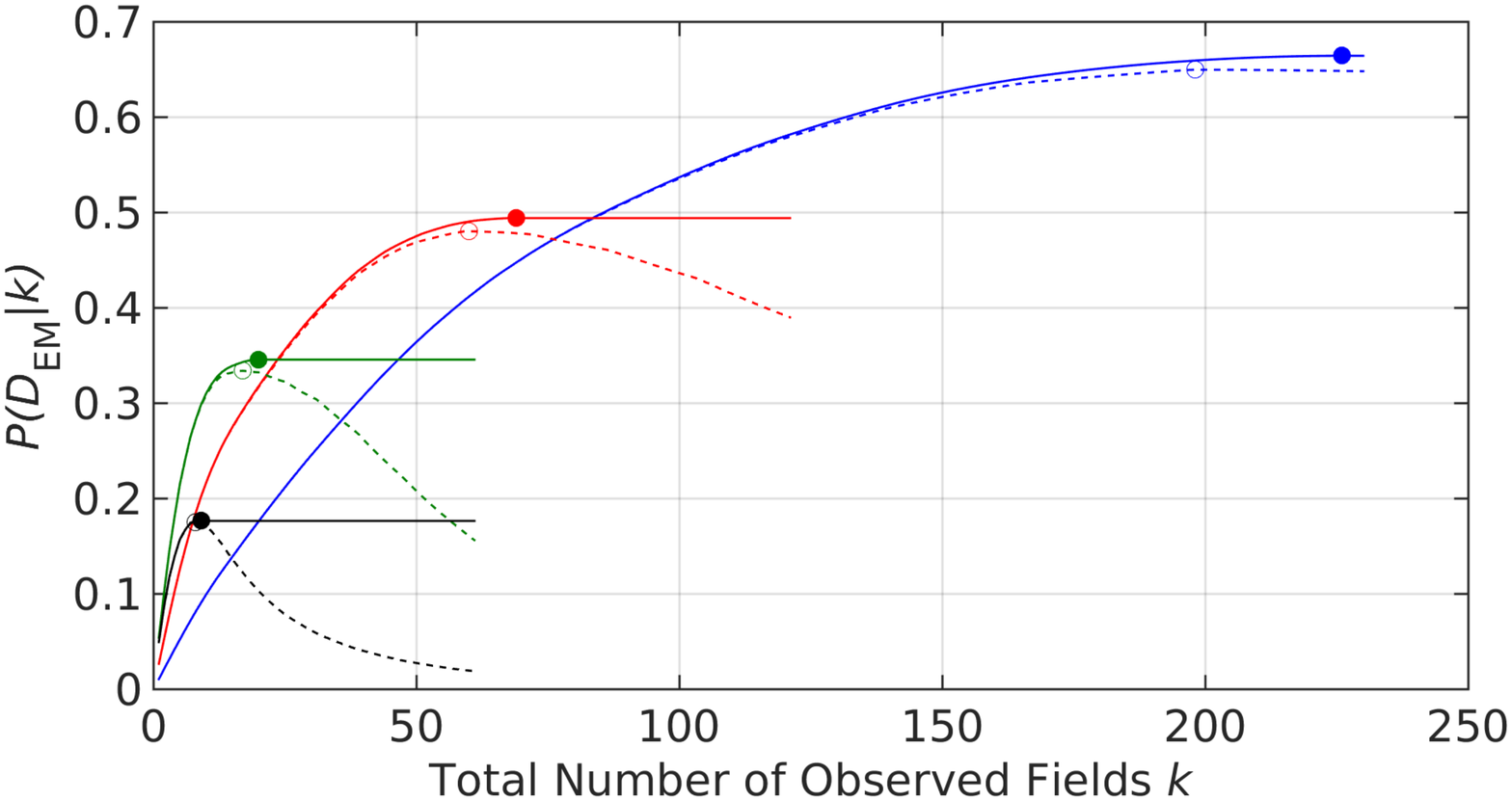}
      \includegraphics[width=.45\textwidth]{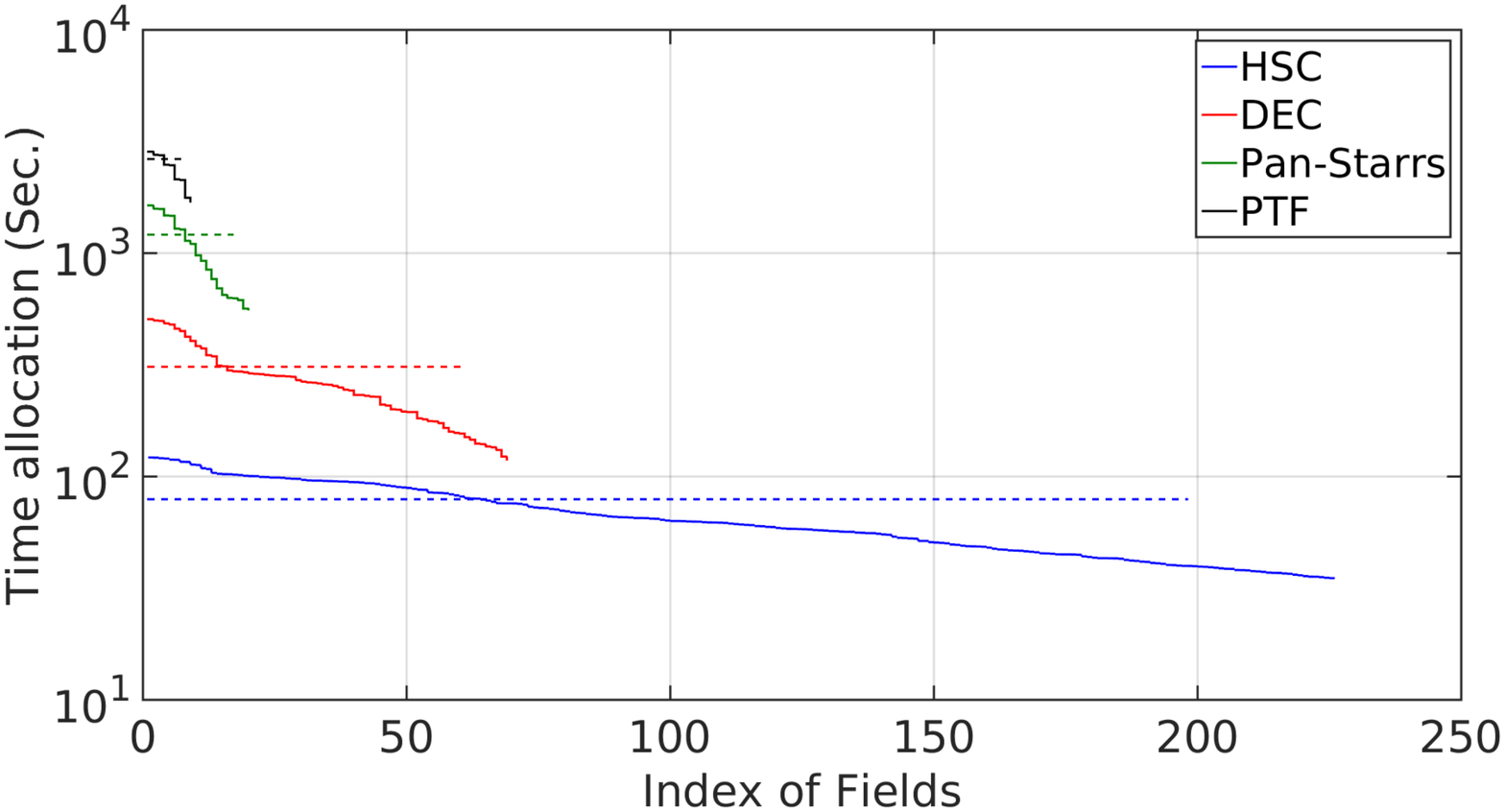}
    }
    \subfigure[\ac{EM} detection probability and optimized observation
times for $T=4$ hour total observation times.]
    {
      \includegraphics[width=.45\textwidth]{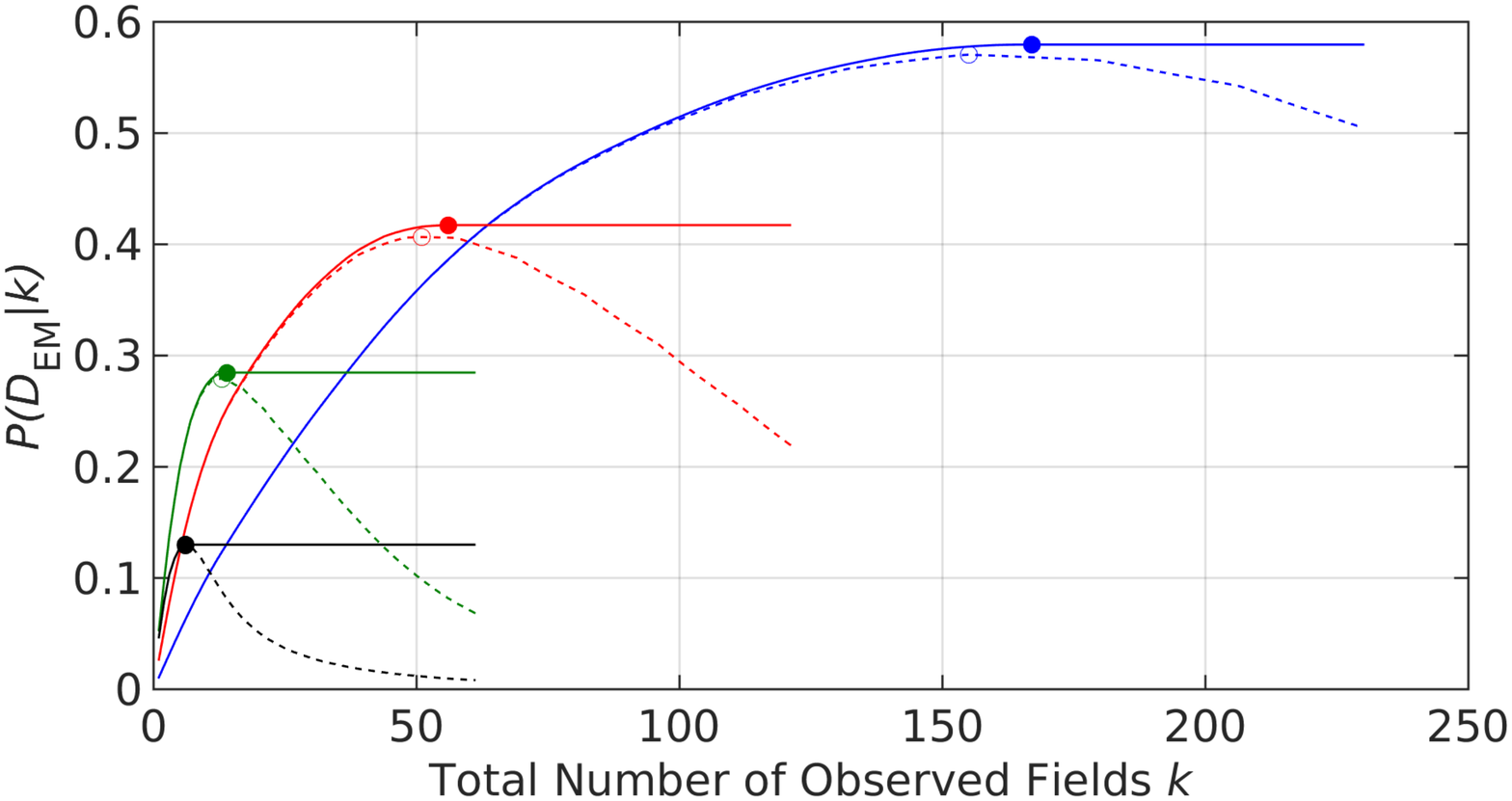}
      \includegraphics[width=.45\textwidth]{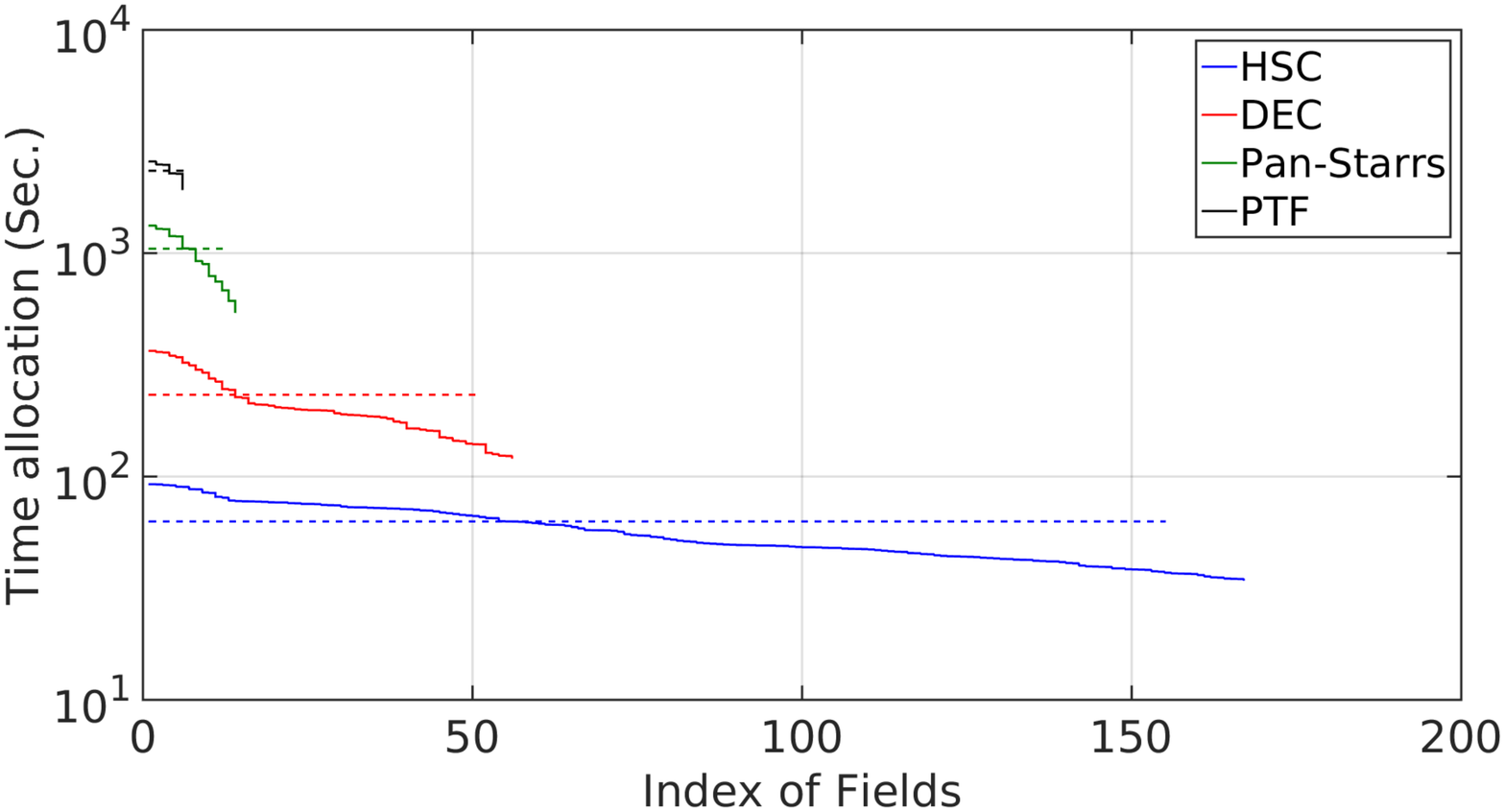}
    }
    \subfigure[\ac{EM} detection probability and optimized observation
times for $T=2$ hour total observation times.]
    {
      \includegraphics[width=.45\textwidth]{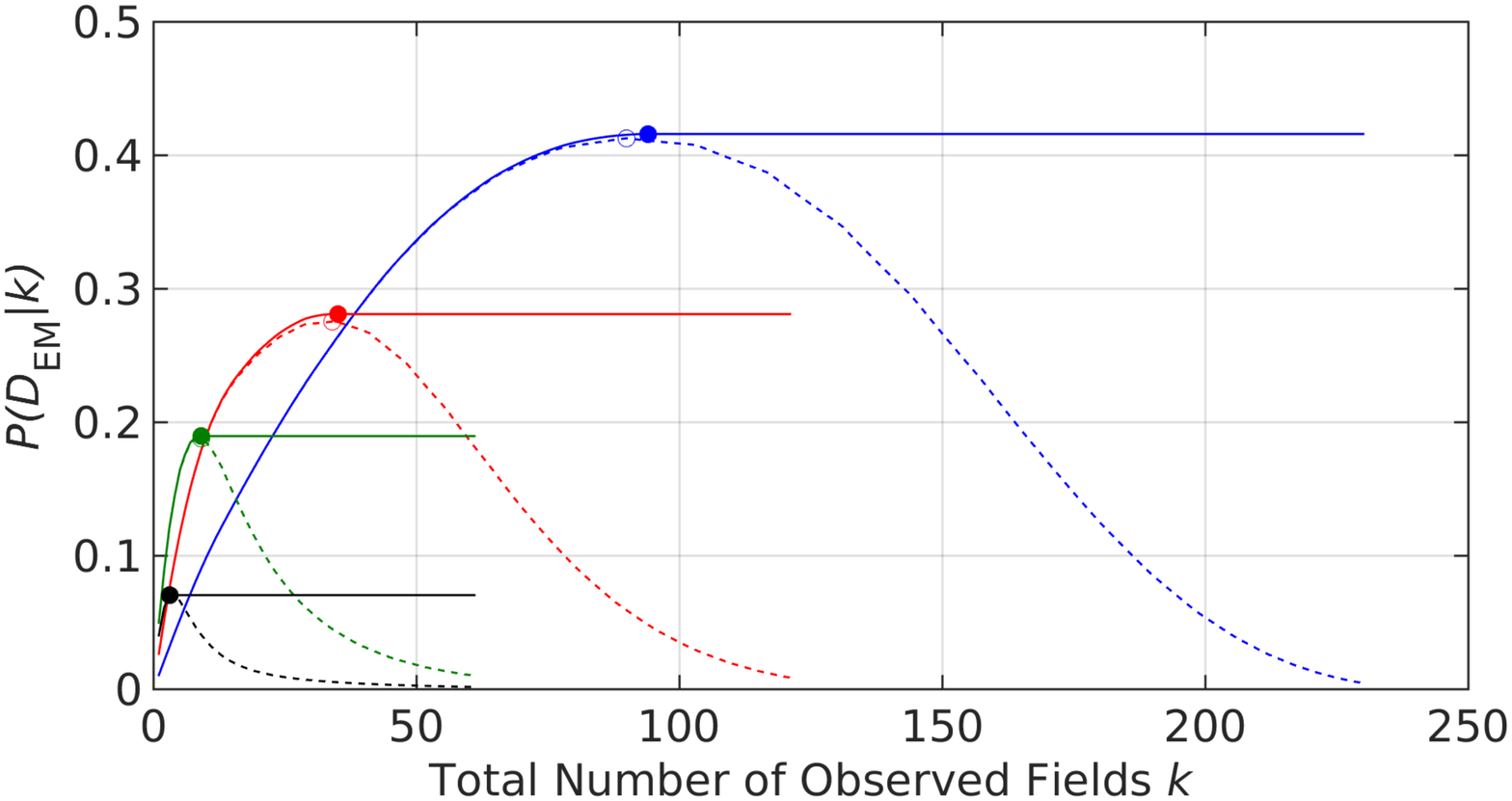}
      \includegraphics[width=.45\textwidth]{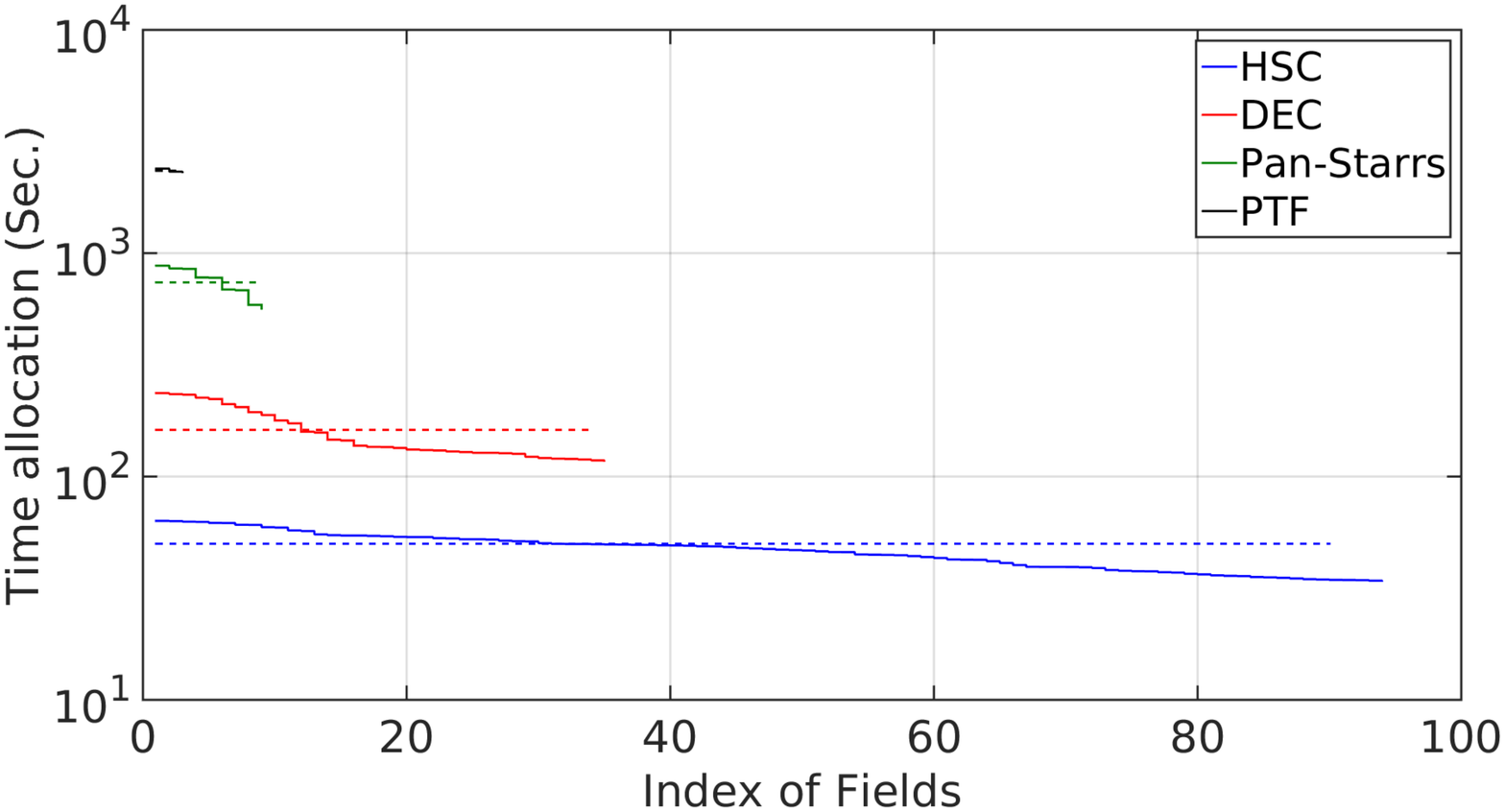}
    }
   \caption{The results of simulated \ac{EM} follow-up observations for the
${\sim}300\,\text{deg}^{2}$ \ac{GW} event (ID 28700). We show the optimized
\ac{EM} detection probability as a function of the number of observing fields
(left) and the allocated observing times for the optimal number of fields
(right). The subfigures (a), (b) and (c) show results from 3 different total 
observation times for 6~hrs, 4~hrs and 2~hrs, respectively. For each total 
observation time the 4 solid curves 
in each plot correspond to the optimal time allocation strategy applied to each
of the 4 telescopes. The dashed lines show results for the equal time strategy. 
The solid markers and the circles indicate the number of observing fields at 
which the maximum detection probability is achieved using the optimal time 
allocation strategy and the equal time strategy respectively.\label{fig:re28700}}
\end{figure*}

\begin{figure*}
  \centering
    \subfigure[\ac{EM} detection probability and optimized observation
times for $T=6$ hour total observation times.]
    {
      \includegraphics[width=.45\textwidth]{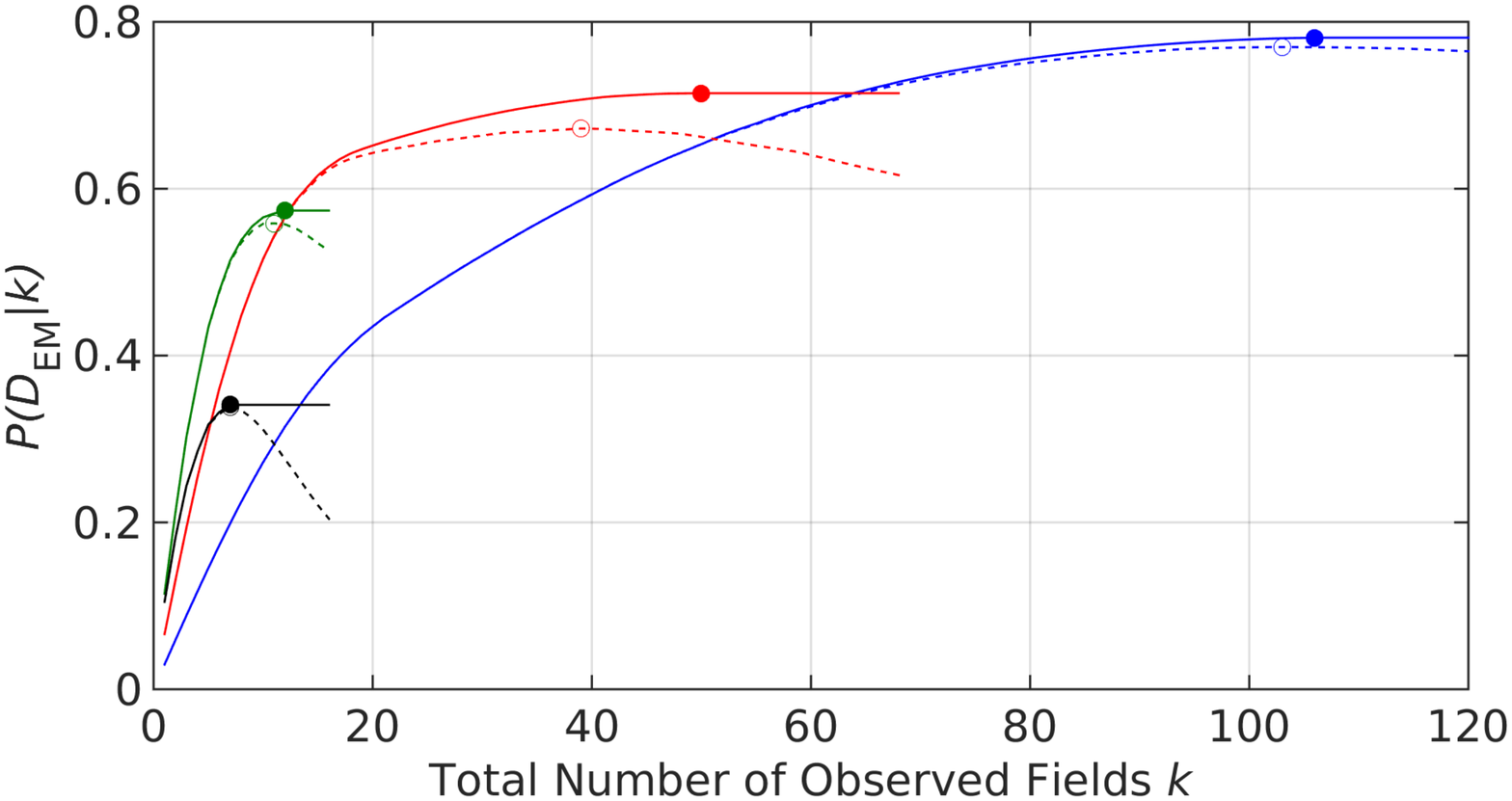}
      \includegraphics[width=.45\textwidth]{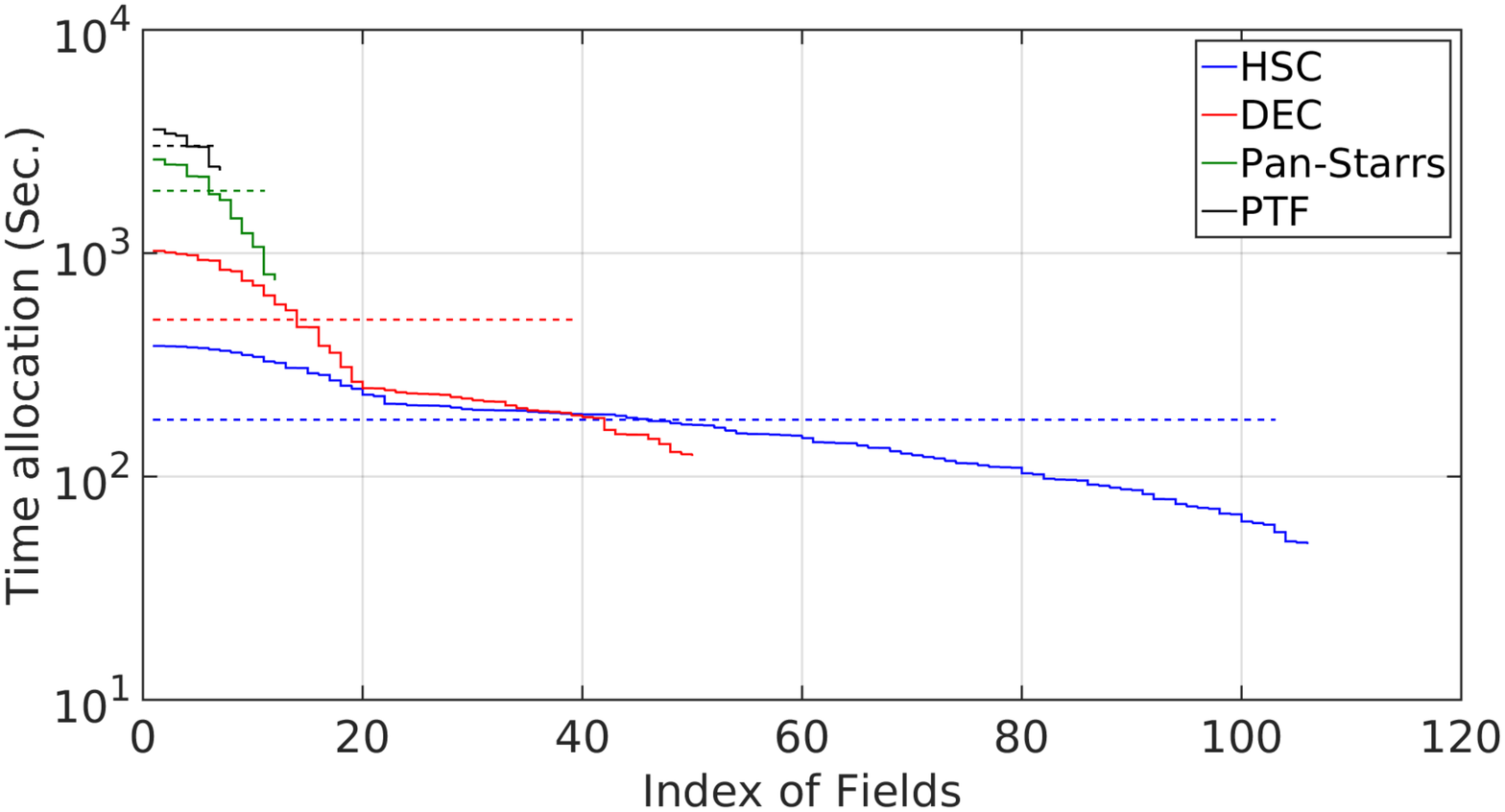}
    }
    \subfigure[\ac{EM} detection probability and optimized observation
times for $T=4$ hour total observation times.]
    {
      \includegraphics[width=.45\textwidth]{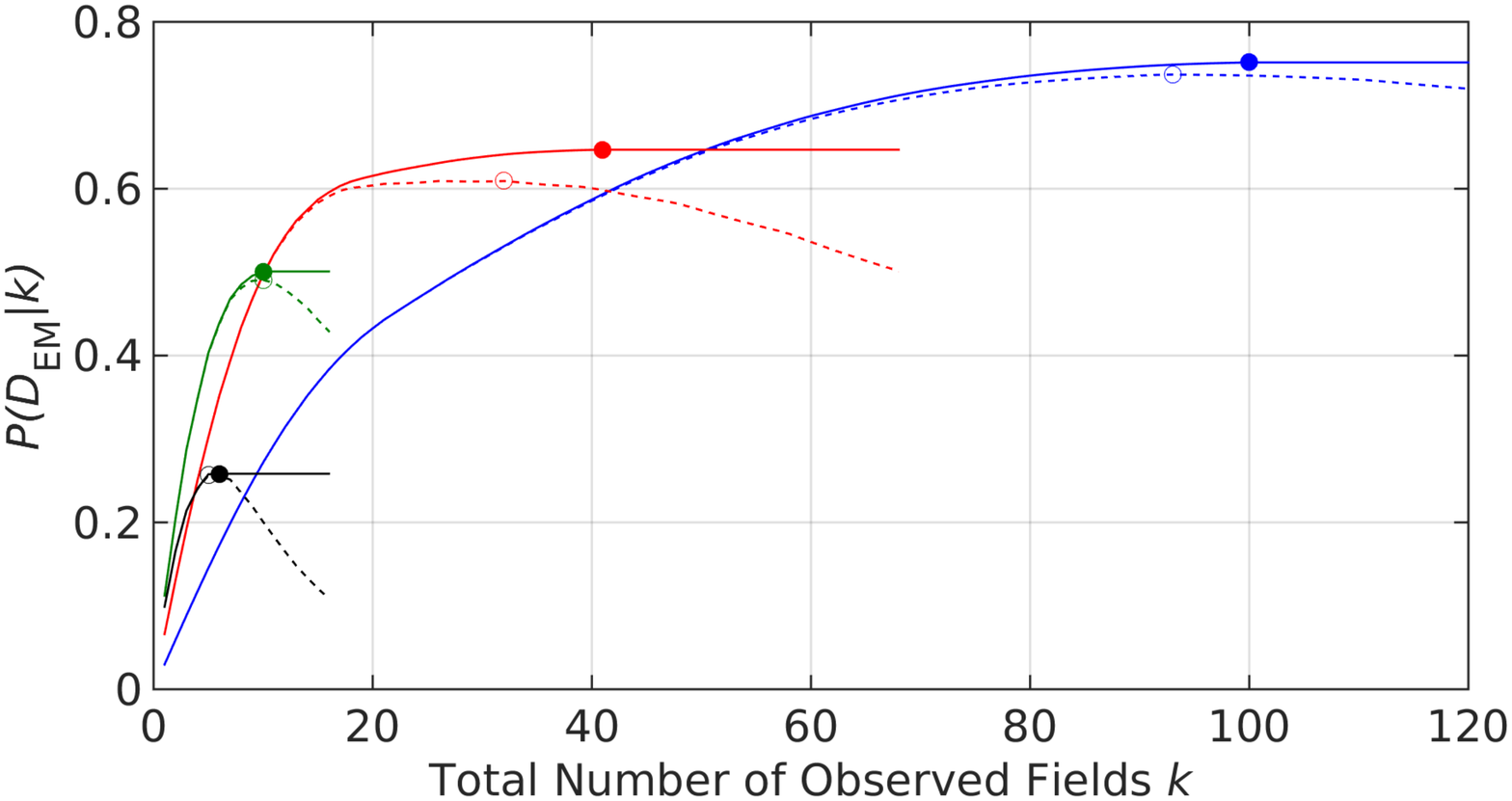}
      \includegraphics[width=.45\textwidth]{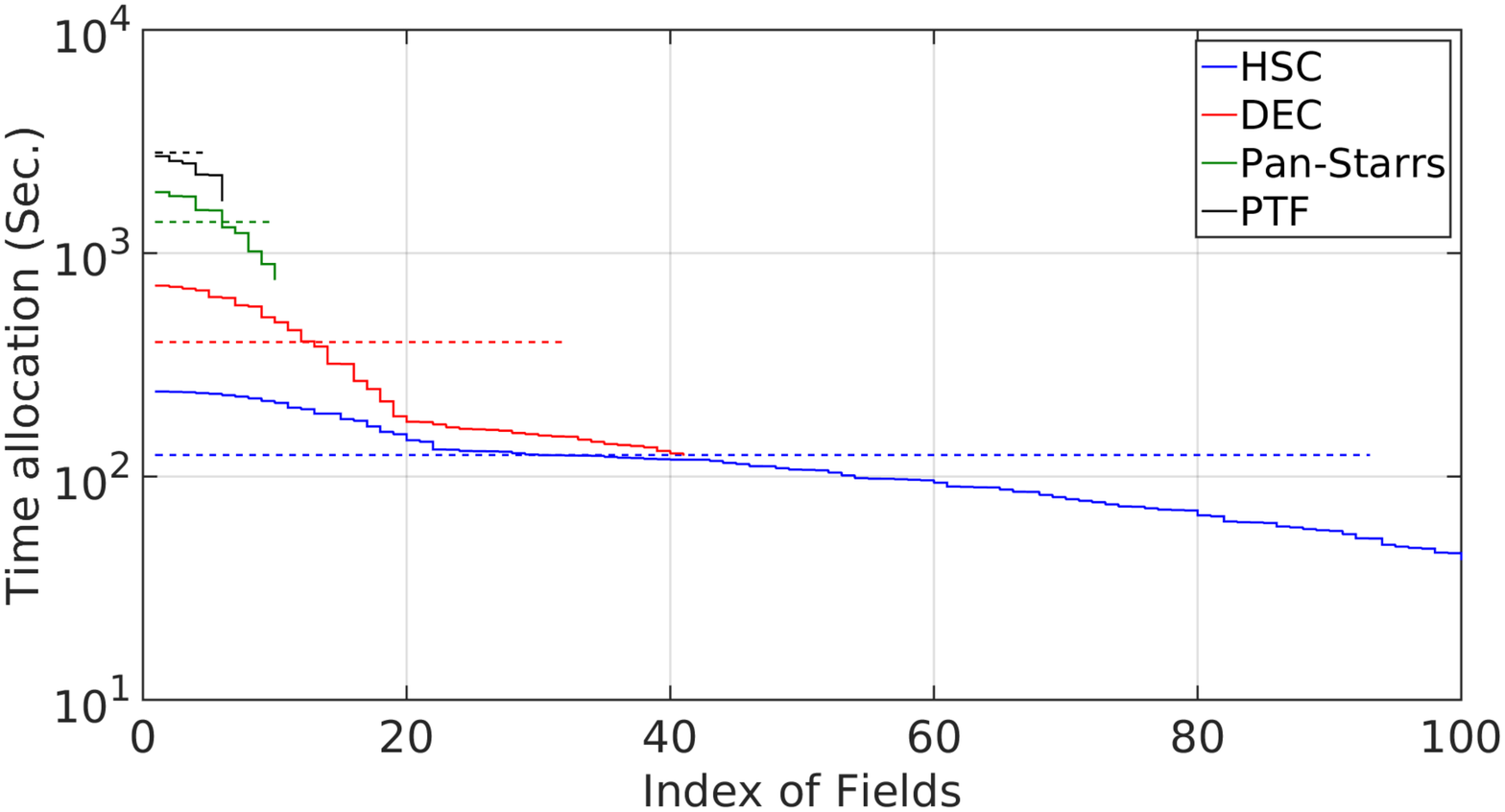}
    }
    \subfigure[\ac{EM} detection probability and optimized observation
times for $T=2$ hour total observation times.]
    {
      \includegraphics[width=.45\textwidth]{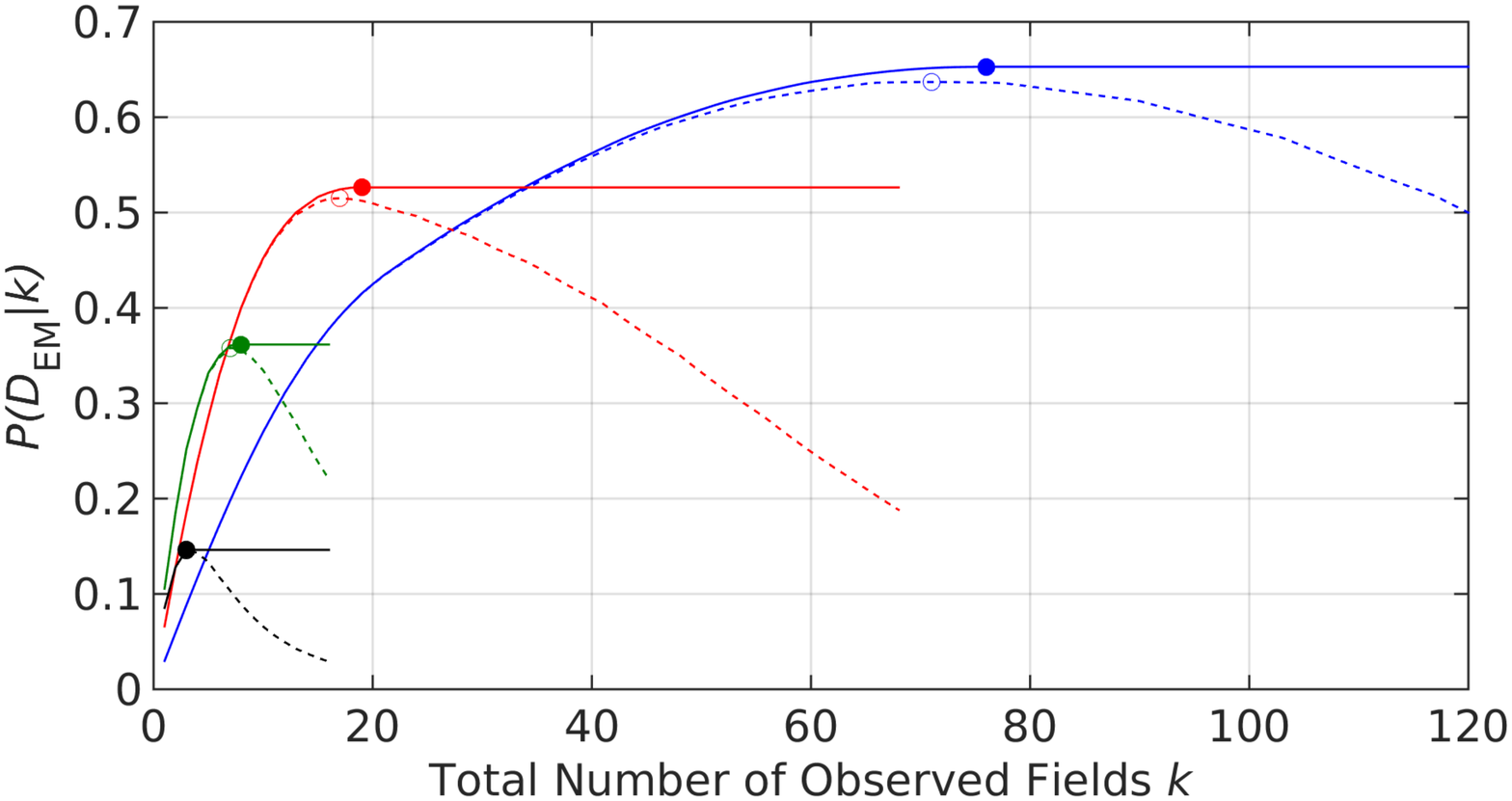}
      \includegraphics[width=.45\textwidth]{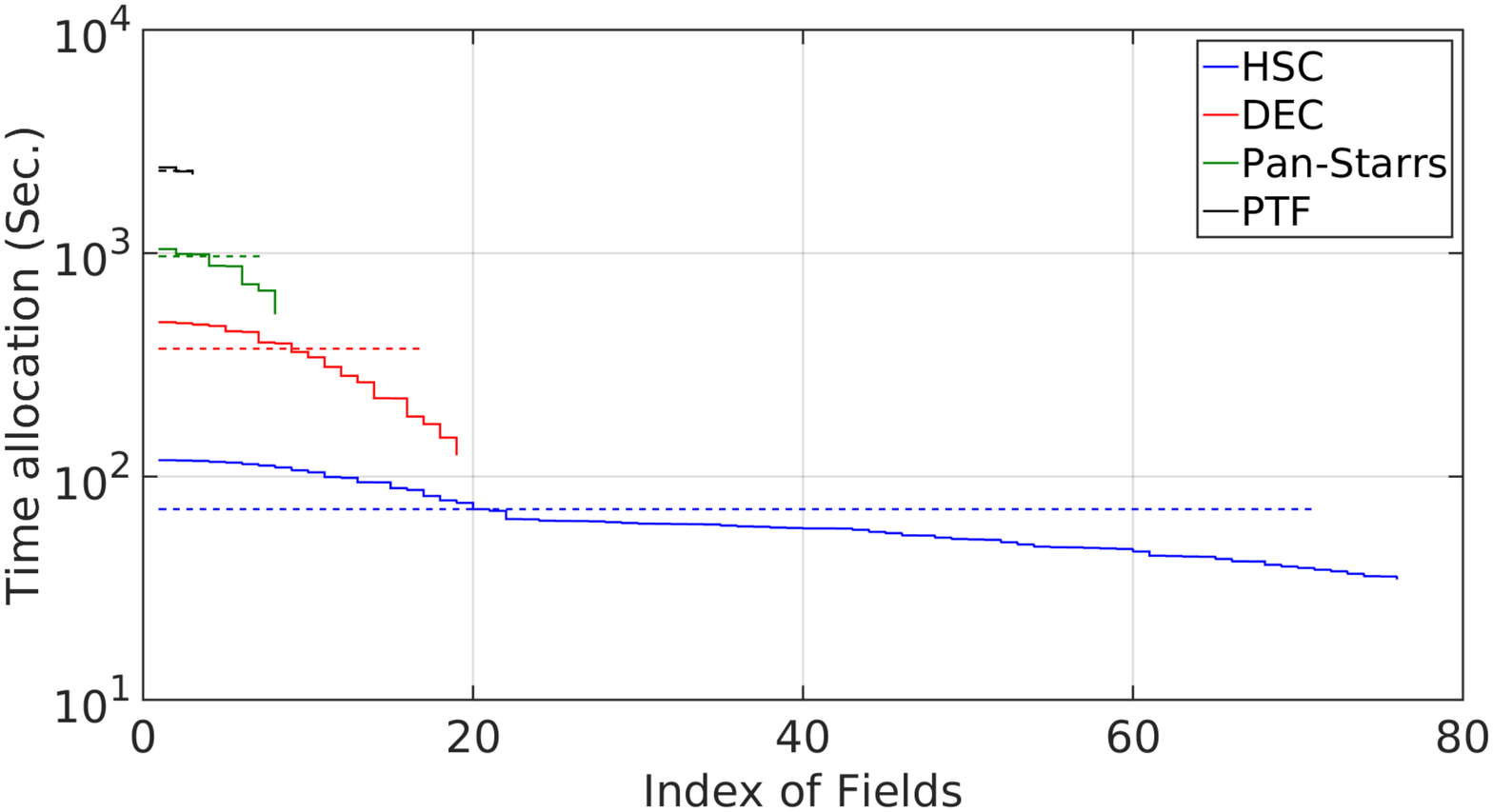}
    }
    \caption{The results of simulated \ac{EM} follow-up observations for the
${\sim}100\,\text{deg}^{2}$ \ac{GW} event (ID 19296). We show the optimized
\ac{EM} detection probability as a function of the number of observing fields
(left) and the allocated observing times for the optimal number of fields
(right). The subfigures (a), (b) and (c) show results from 3 different total 
observation times for 6~hrs, 4~hrs and 2~hrs, respectively. 
For each total observation time the 4 solid curves
in each plot correspond to the optimal time allocation strategy applied to each
of the 4 telescopes. The dashed lines show results for the equal time strategy.
The solid markers and the circles indicate the number of observing fields at 
which the maximum detection probability is achieved using the optimal time allocation
strategy and the equal time strategy respectively.\label{fig:re19296}}
\end{figure*}

\begin{figure*}
  \centering
    \subfigure[\ac{EM} detection probability and optimized observation
times for $T=6$ hour total observation times.]
    {
      \includegraphics[width=.45\textwidth]{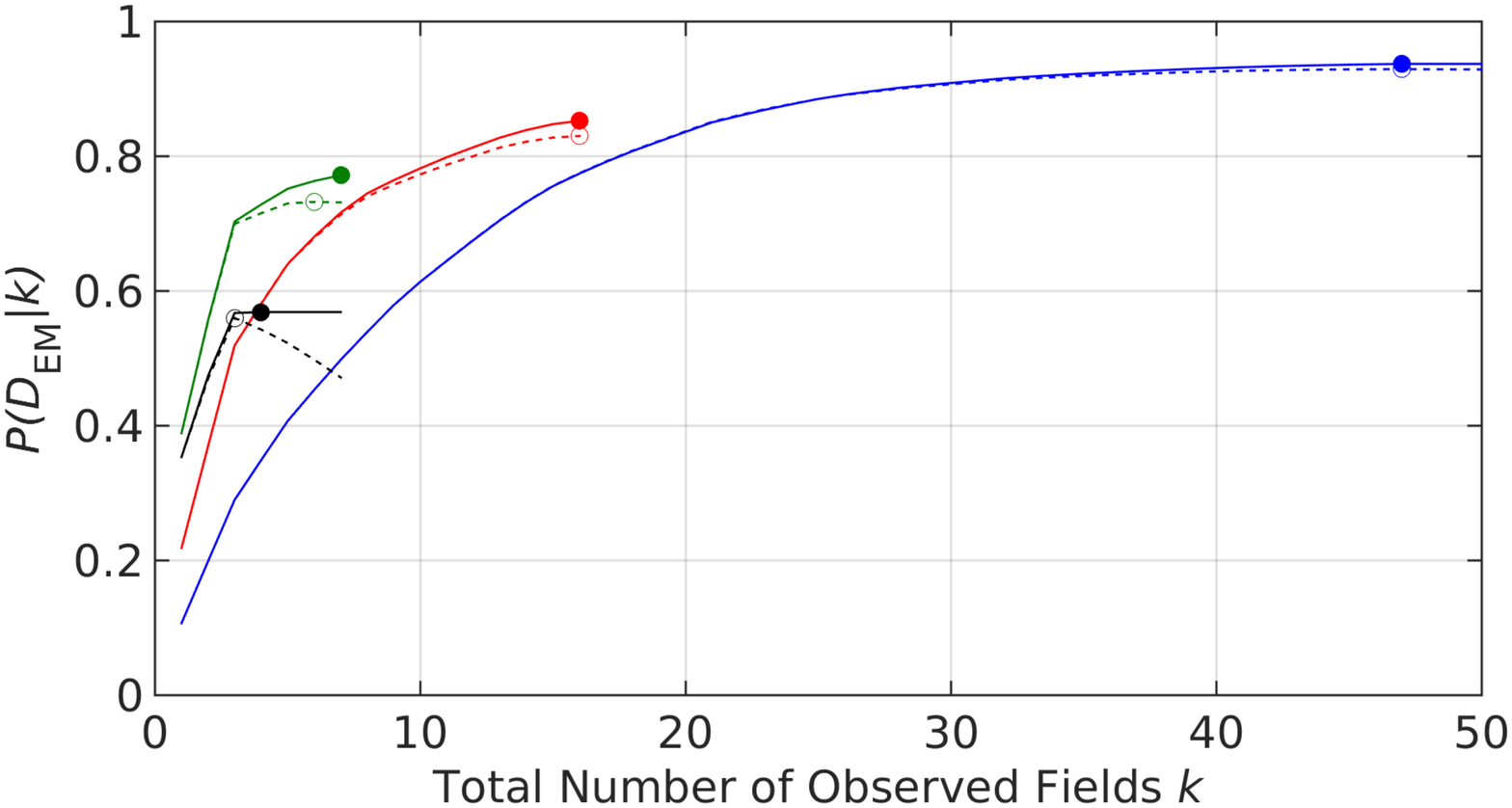}
      \includegraphics[width=.45\textwidth]{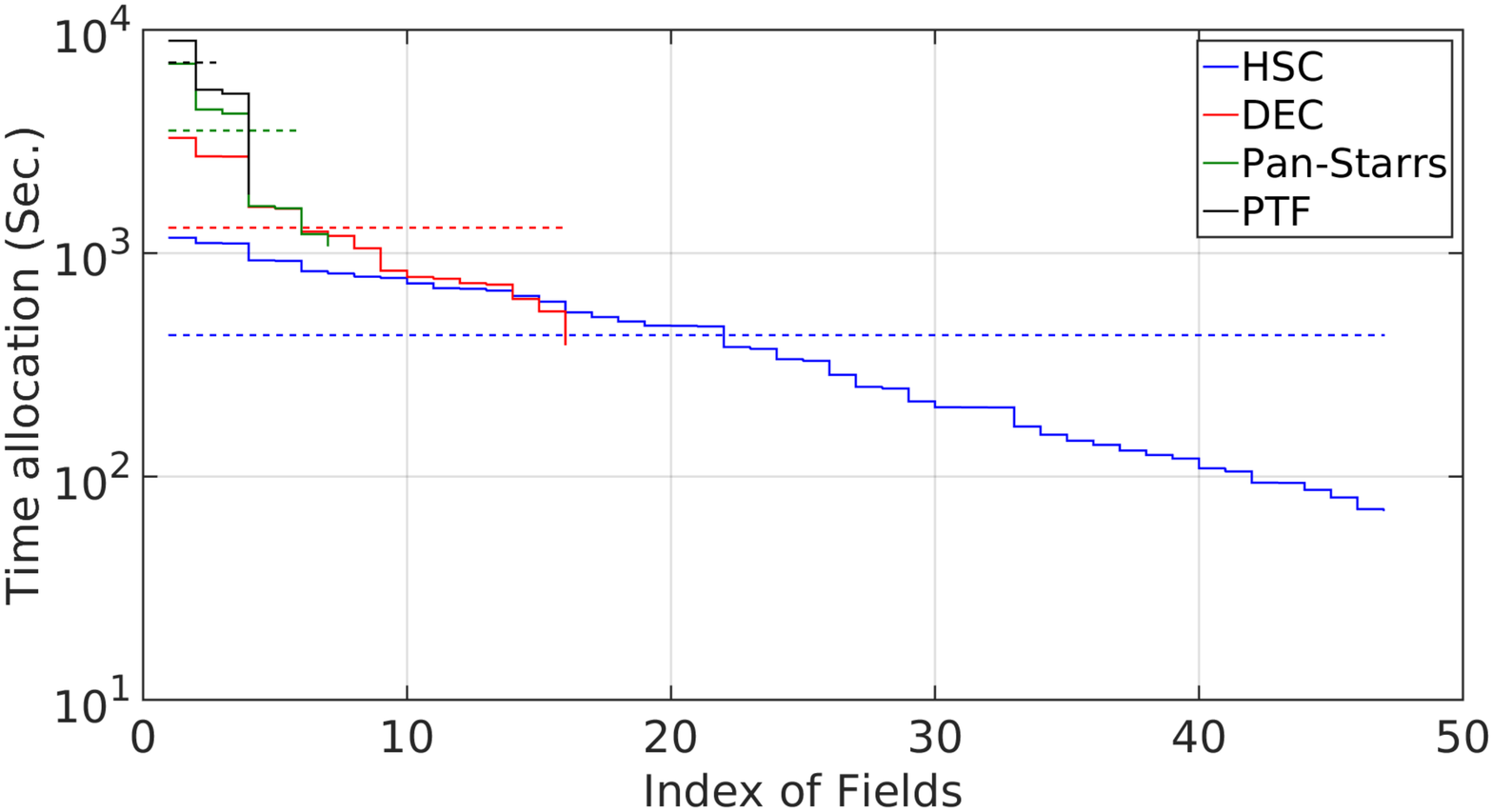}
    }
    \subfigure[\ac{EM} detection probability and optimized observation
times for $T=4$ hour total observation times.]
    {
      \includegraphics[width=.45\textwidth]{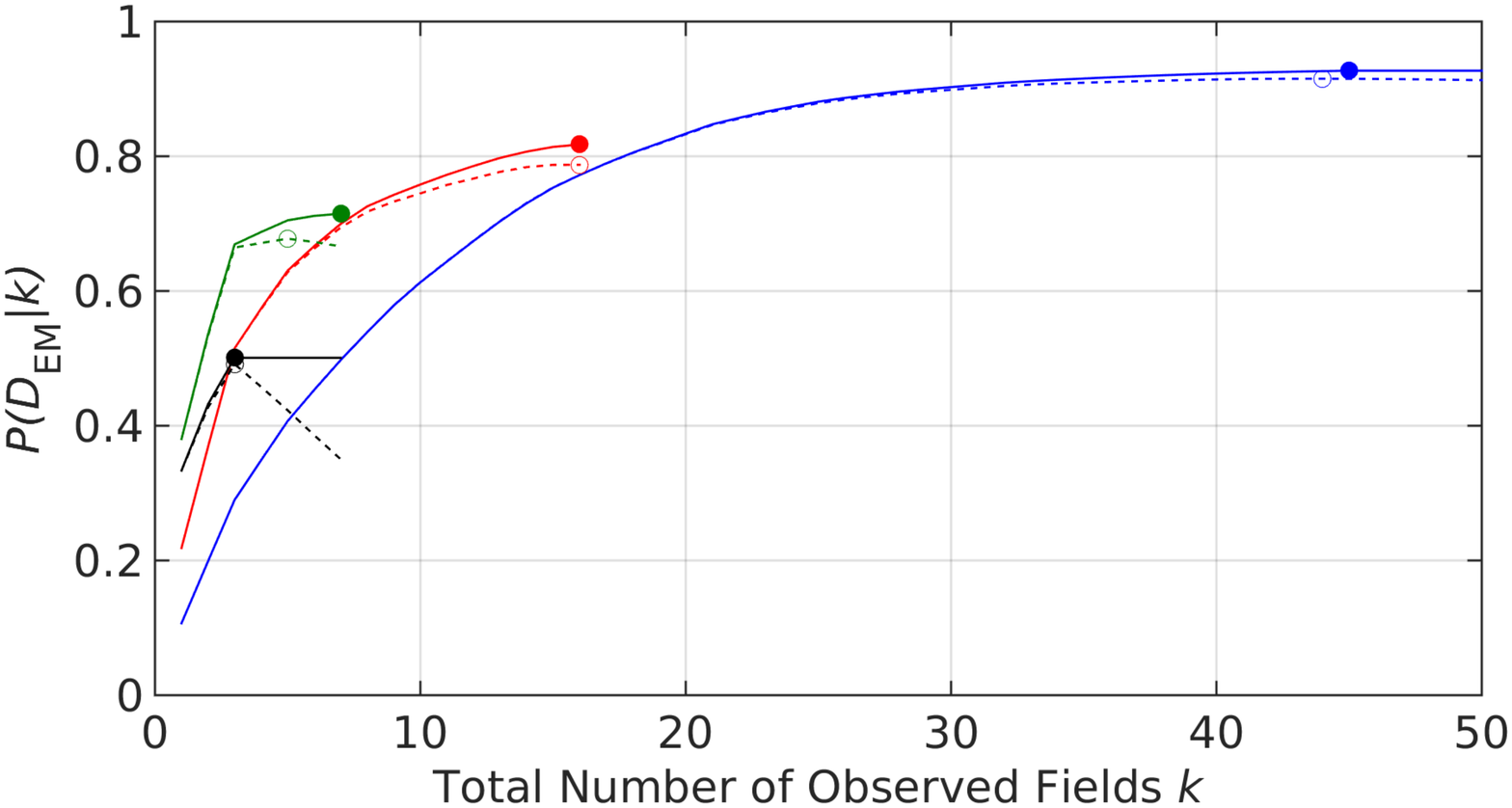}
      \includegraphics[width=.45\textwidth]{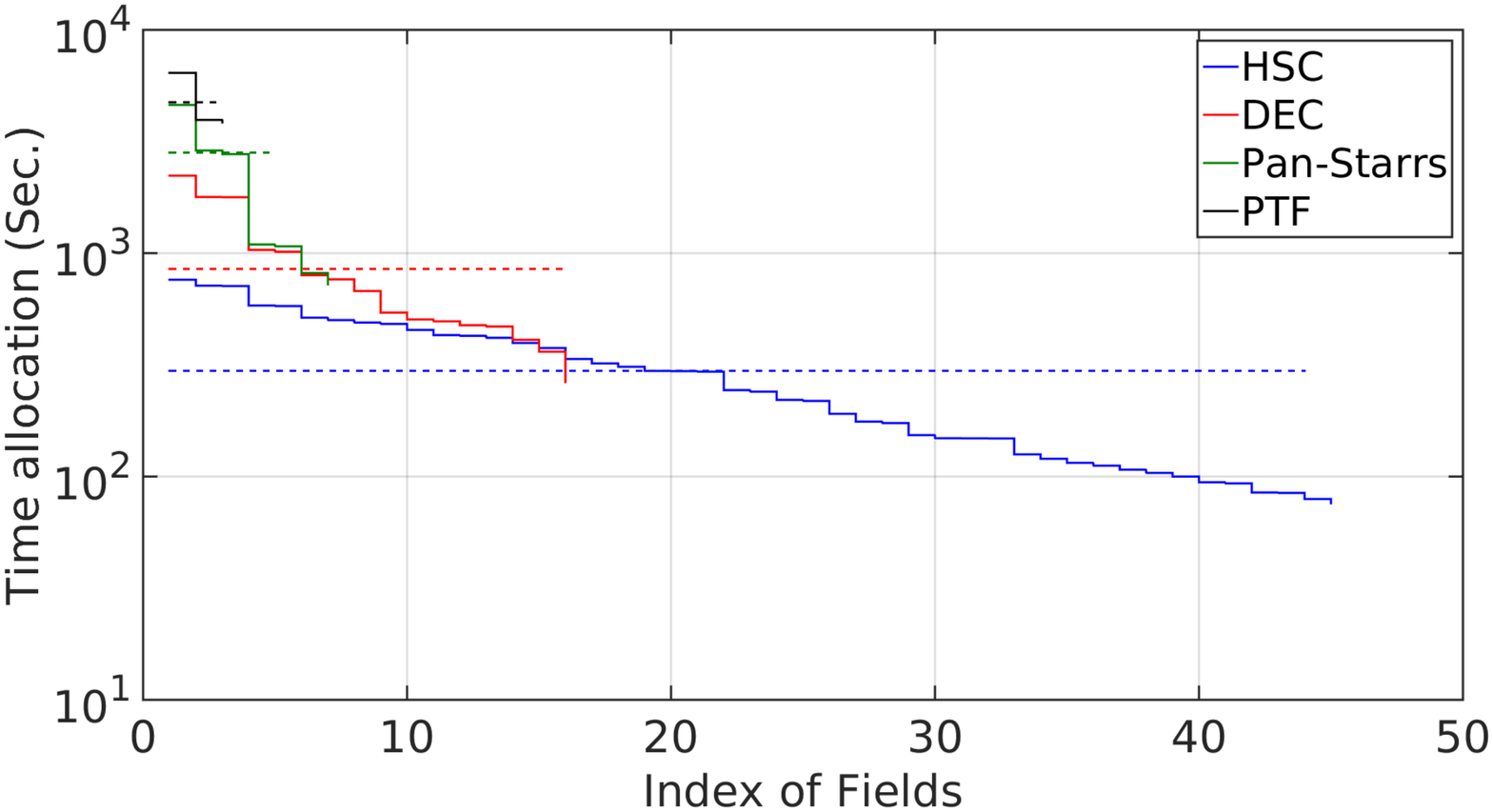}
    }
    \subfigure[\ac{EM} detection probability and optimized observation
times for $T=2$ hour total observation times.]
    {
      \includegraphics[width=.45\textwidth]{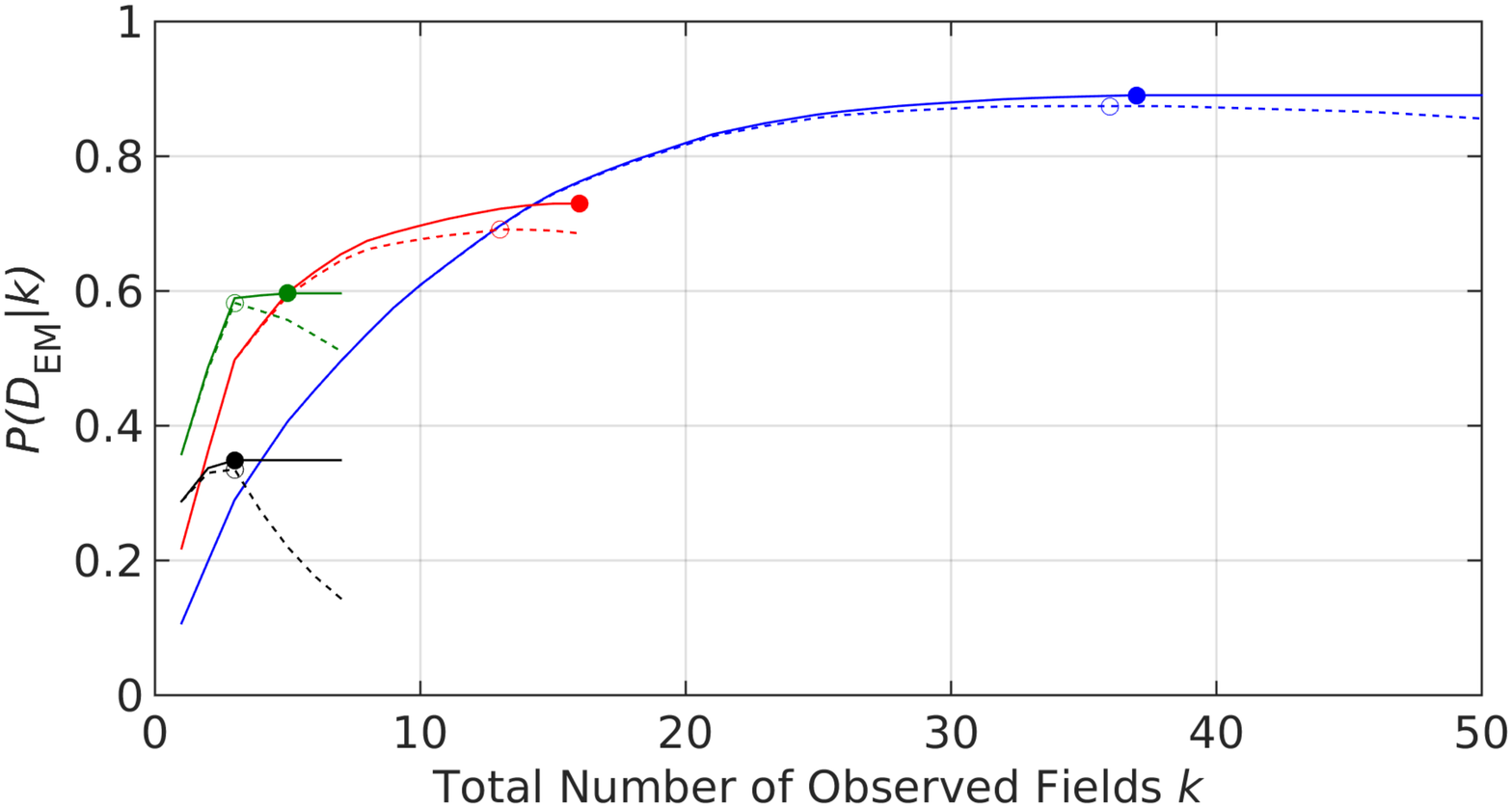}
      \includegraphics[width=.45\textwidth]{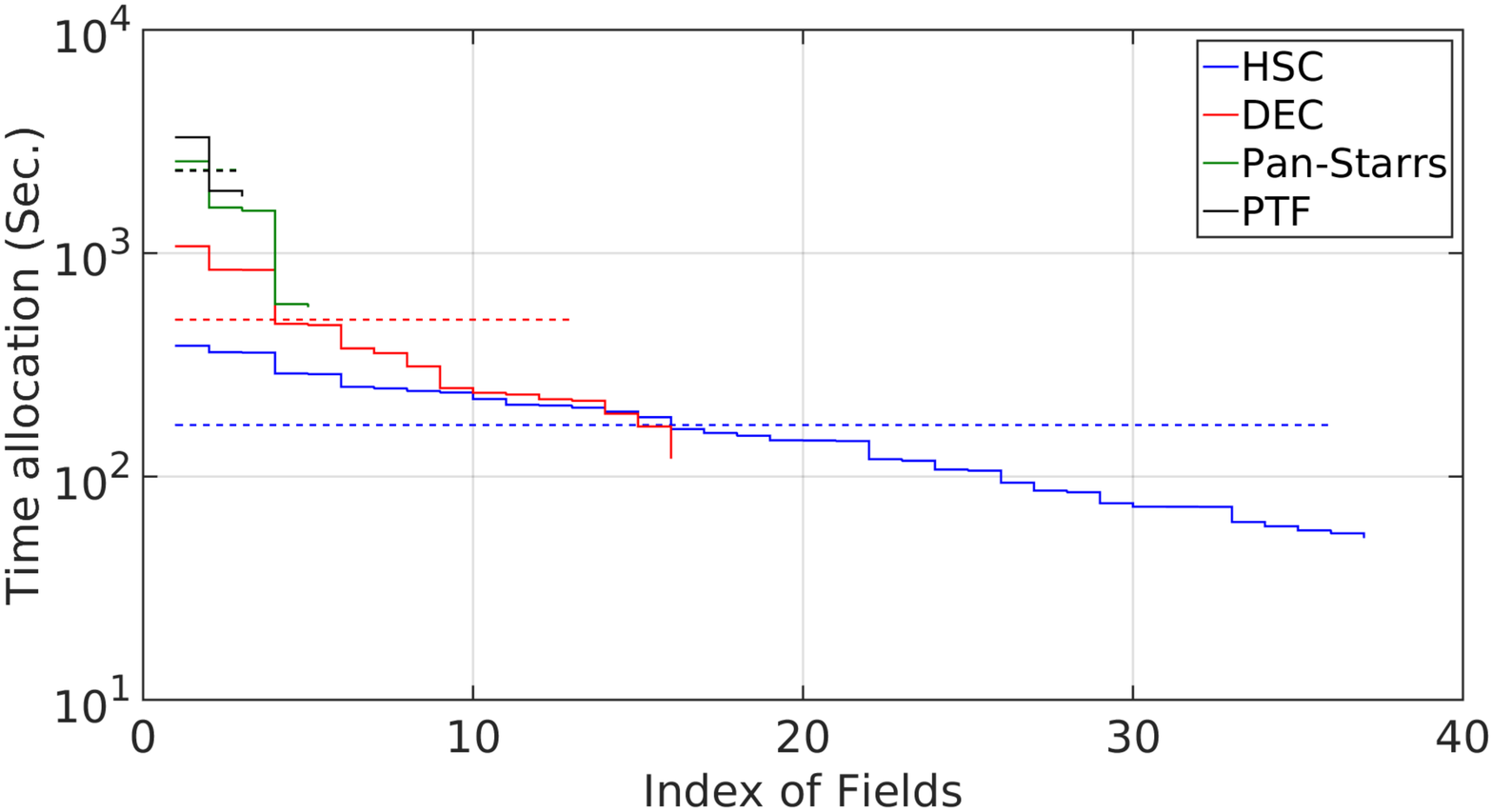}
    }
    \caption{The results of simulated \ac{EM} follow-up observations for the
${\sim}30\,\text{deg}^{2}$ \ac{GW} event (ID 18694). We show the optimized
\ac{EM} detection probability as a function of the number of observing fields
(left) and the allocated observing times for the optimal number of fields
(right). The subfigures (a), (b) and (c) show results from 3 different total 
observation times for 6~hrs, 4~hrs and 2~hrs, respectively. 
For each total observation time the 4 solid curves
in each plot correspond to the optimal time allocation strategy applied to each
of the 4 telescopes. The dashed lines show results for the equal time strategy.
The solid markers and the circles indicate the number of observing fields at 
which the maximum detection probability is achieved using the optimal time 
allocation strategy and the equal time strategy respectively.\label{fig:re18694}}
\end{figure*}

%
Figures~\ref{fig:re28700},~\ref{fig:re19296} and~\ref{fig:re18694} display the
results of the simulated \ac{EM} follow-up observations for the events labeled
28700, 19296 and 18694 respectively. From the top panels to the bottom panels,
the assumed total observation times are 6\,hrs, 4\,hrs and 2\,hrs. The plots on
the left of the figure show the optimized detection probability
$P(D_{\mathrm{EM}}|k)$ as a function of the total observed number of fields
$k$. The plots on the right of the figure display the optimal time allocations
corresponding to the value of $k$ returning the highest detection probability
(indicated by a circular marker in the detection probability plots on the left
of the figure). The indices refer to the labels assigned to the fields when
being chosen by the greedy algorithm.

%
We highlight the asymptotic behavior of the detection probability curves in
Figs.~\ref{fig:re28700},~\ref{fig:re19296} and~\ref{fig:re18694}. This is due
to a particular feature of our algorithm and is explained as follows.  As the
number of fields are increased we approach an optimal value $k^{*}$ where to
observe an addition field with any finite observation time would actually
reduce the detection probability. This occurs where the gains from an
additional field are outweighed by the losses incurred by reducing the lengths
of the observations of the other fields. In this case, for a given value of $k$
above the optimal value the optimal choice is to allocate $\tau=0$ to all
fields with index greater than $k^{*}$. If we know that we will allocate no
time to these fields then we also have no need to slew to them or to readout
from the \ac{CCD}. Hence the optimal time allocations and also detection
probability for values of $k>k^{*}$ remains constant at the maximum value.

%
%
%
%
%

%
As a comparison, we introduce a second observing strategy for time allocation
in which all the fields are observed with equal time. We call this method the
equal time strategy. For each value of $k$, subject to the total observation
time constraints and slew/readout time, the time allocated to each field is
given by $T/k-T_{0}$. The values of $P(D_{\mathrm{EM}}|k)$ obtained using this
strategy are plotted as dashed lines in the plots on the left of
Figs.~\ref{fig:re28700},~\ref{fig:re19296} and~\ref{fig:re18694}. The time
allocations corresponding to the peaks of the dashed lines are plotted as flat
lines (constant equal values) in the plots on the right of the figures. The
resultant maximal probabilities and corresponding optimal number of fields for
both time allocation strategies are given in Table~\ref{table:results} for all
simulated events and total observation times.  We also indicate the relative
gains in detection probability obtained using our fully optimized (Lagrange
multiplier) approach relative to the equal time strategy. 
%
%
\begin{table*}
\centering
  \caption{The \ac{EM} detection probability using both the optimal and equal
time strategies}
    \label{table:results}
  \begin{tabular*}{\textwidth}{@{\extracolsep{\fill}}lccccccccccc}
\hline\hline
Telescope            & Event ID               & Strategy   &
\multicolumn{6}{c}{EM detection probability (optimal number of fields)} & \multicolumn{3}{c}{Relative Gain}                                          \\ 
                     &                        &    & \multicolumn{2}{c}{6\,hrs}
& \multicolumn{2}{c}{4\,hrs}        & \multicolumn{2}{c}{2\,hrs}        & 6 h        & 4 h        & 2 h                                                \\
\hline
\multirow{6}{*}{\ac{HSC}}   & \multirow{2}{*}{28700} & LM\footnote{\scriptsize LM: Lagrange multiplier} & 66.4\% & (226)       & 58.0\% & (167)       & 41.6\% & (94)       & \multirow{2}{*}{1.4\%} &\multirow{2}{*}{1.0\%} &\multirow{2}{*}{0.3\%} \\
                            &                        & ET\footnote{\scriptsize ET: Equal time strategy} & 65.0\% & (198)       & 57.0\% & (155)       & 41.3\% & (90)       &                                                \\

                            & \multirow{2}{*}{19296} & LM & 78.1\% &(106) & 75.1\% & (100)      & 65.3\% & (76)       & \multirow{2}{*}{1.1\%} &\multirow{2}{*}{1.4\%} &\multirow{2}{*}{1.6\%} \\
                            &                        & ET & 77.0\% &(103) & 73.7\% & (93)       & 63.7\% & (71)       &                                                \\
                            & \multirow{2}{*}{18694} & LM & 93.7\% &(47)  & 92.7\% & (45)       & 89.1\% & (37)       & \multirow{2}{*}{0.8\%} &\multirow{2}{*}{1.2\%} &\multirow{2}{*}{1.7\%} \\
                            &                        & ET & 92.9\% &(47)  & 91.5\% & (44)       & 87.4\% & (36)       &                                                \\
\hline
\multirow{6}{*}{\ac{DEC}}   & \multirow{2}{*}{28700} & LM & 49.4\% &(69)  & 41.7\% & (56)       & 28.1\% & (35)       & \multirow{2}{*}{1.4\%} &\multirow{2}{*}{1.0\%} &\multirow{2}{*}{0.5\%} \\
		            &                        & ET & 48.0\% &(60)  & 40.7\% & (51)       & 27.6\% & (34)       &                                            \\
                            & \multirow{2}{*}{19296} & LM & 71.5\% &(50)  & 64.7\% & (41)       & 52.6\% & (19)       & \multirow{2}{*}{4.3\%} &\multirow{2}{*}{3.8\%} &\multirow{2}{*}{1.1\%} \\
                            &                        & ET & 67.2\% &(39)  & 60.9\% & (32)       & 51.5\% & (17)       &                                                \\
                            & \multirow{2}{*}{18694} & LM & 85.3\% &(16)  & 81.7\% & (16)       & 73.0\% & (16)       & \multirow{2}{*}{2.3\%} &\multirow{2}{*}{3.0\%} &\multirow{2}{*}{3.9\%} \\
                            &                        & ET & 83.0\% &(16)  & 78.7\% & (16)       & 69.1\% & (13)       &                                                \\
                            
\hline
\multirow{6}{*}{Pan-Starrs} & \multirow{2}{*}{28700} & LM & 34.6\% &(20)  & 28.5\% & (14)       & 19.0\% & (9)       & \multirow{2}{*}{1.2\%} &\multirow{2}{*}{0.5\%} &\multirow{2}{*}{0.2\%} \\
                            &                        & ET & 33.4\% &(17)  & 28.0\% & (13)       & 18.8\% & (9)       &                                                \\
                            & \multirow{2}{*}{19296} & LM & 57.4\% &(12)  & 50.1\% & (10)       & 36.2\% & (8)       & \multirow{2}{*}{1.5\%} &\multirow{2}{*}{1.0\%} &\multirow{2}{*}{0.4\%} \\
                            &                        & ET & 55.9\% &(11)  & 49.1\% & (10)       & 35.8\% & (7)       &                                                \\
                            & \multirow{2}{*}{18694} & LM & 77.2\% &(7)   & 71.5\% & (7)       & 59.7\% & (5)       & \multirow{2}{*}{4.0\%} &\multirow{2}{*}{3.7\%} &\multirow{2}{*}{1.4\%} \\
                            &                        & ET & 73.2\% &(6)   & 67.8\% & (5)       & 58.3\% & (3)       &                                                \\
\hline
\multirow{6}{*}{\ac{PTF}}   & \multirow{2}{*}{28700} & LM & 17.7\% &(9)   & 13.0\% & (6)       & 7.1\% & (3)       & \multirow{2}{*}{0.2\%} &\multirow{2}{*}{0.1\%} &\multirow{2}{*}{0.0\%} \\
                            &                        & ET & 17.5\% &(8)   & 12.9\% & (6)       & 7.1\% & (3)       &                                                \\
                            & \multirow{2}{*}{19296} & LM & 34.1\% &(7)   & 25.8\% & (6)       & 14.6\% & (3)       & \multirow{2}{*}{0.3\%} &\multirow{2}{*}{0.2\%} &\multirow{2}{*}{0.0\%} \\
                            &                        & ET & 33.8\% &(7)   & 25.6\% & (5)       & 14.6\% & (3)       &                                                \\
                            & \multirow{2}{*}{18694} & LM & 56.9\% &(4)   & 50.1\% & (3)       & 34.9\% & (3)       & \multirow{2}{*}{0.8\%} &\multirow{2}{*}{0.9\%} &\multirow{2}{*}{1.4\%} \\ 
                            &                        & ET & 56.1\% &(3)   & 49.2\% & (3)       & 33.5\% & (3)       &                                                \\
\hline\hline  
\end{tabular*}
\end{table*}	

\begin{figure*}
  \centering
    \subfigure[The \ac{EM} detection probability and optimized number of fields
for a general telescope observing event 28700 (${\sim}300\,\text{deg}^{2}$).]
    {
      \includegraphics[width=.45\textwidth]{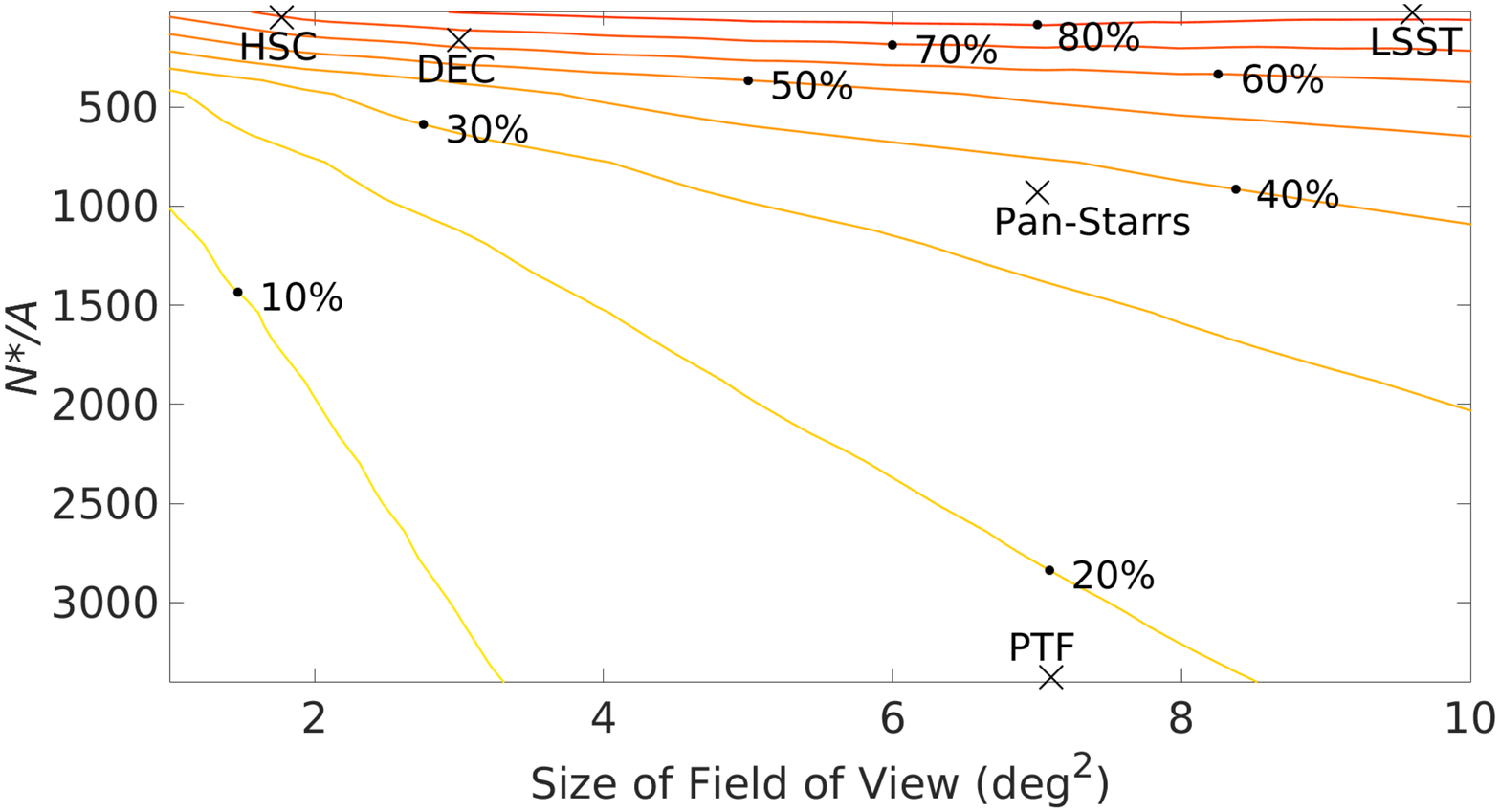}
      \includegraphics[width=.45\textwidth]{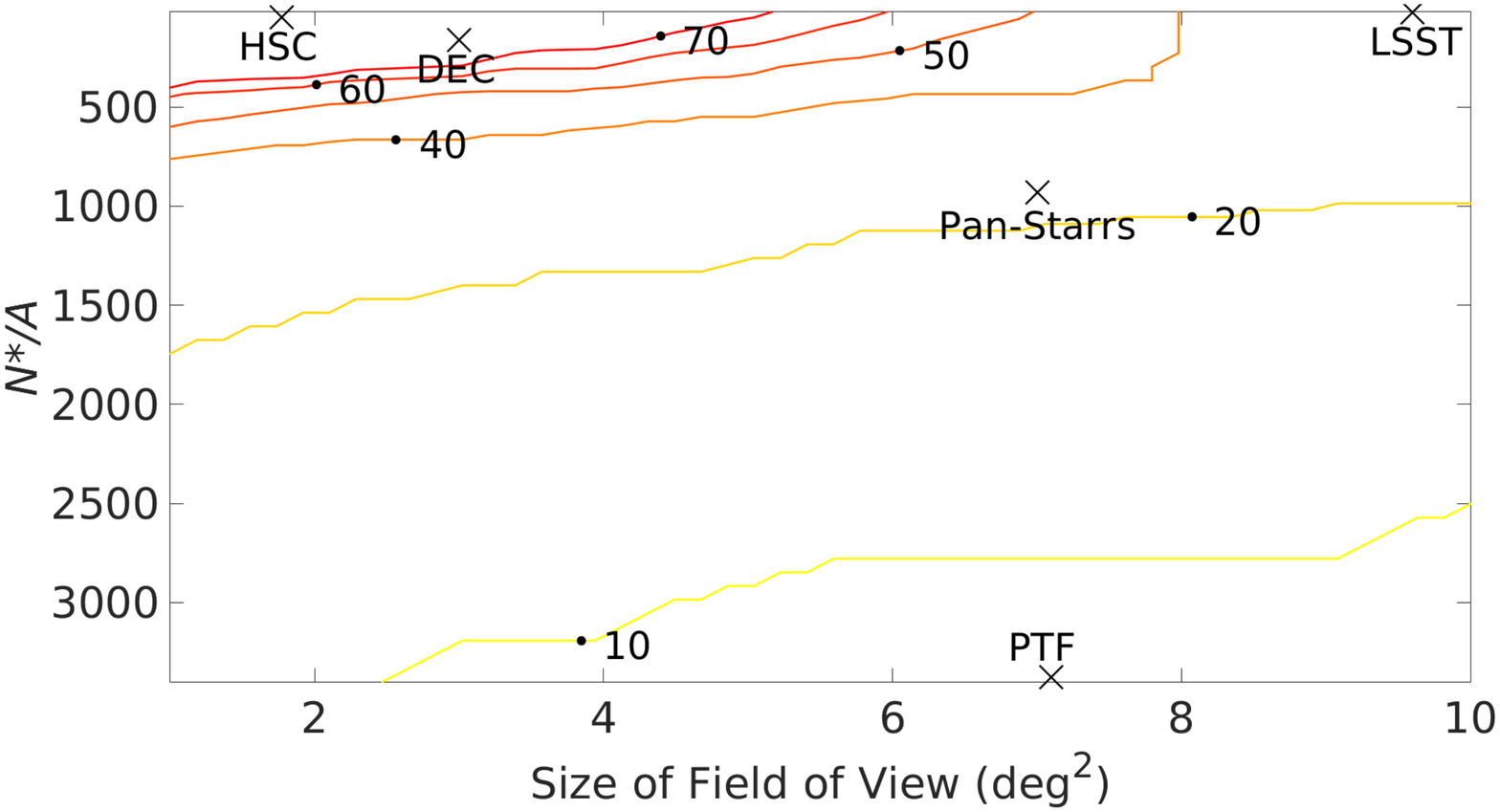}
    }
    \subfigure[The \ac{EM} detection probability and optimized number of fields
for a general telescope observing event 19296 (${\sim}100\,\text{deg}^{2}$).]
    {
      \includegraphics[width=.45\textwidth]{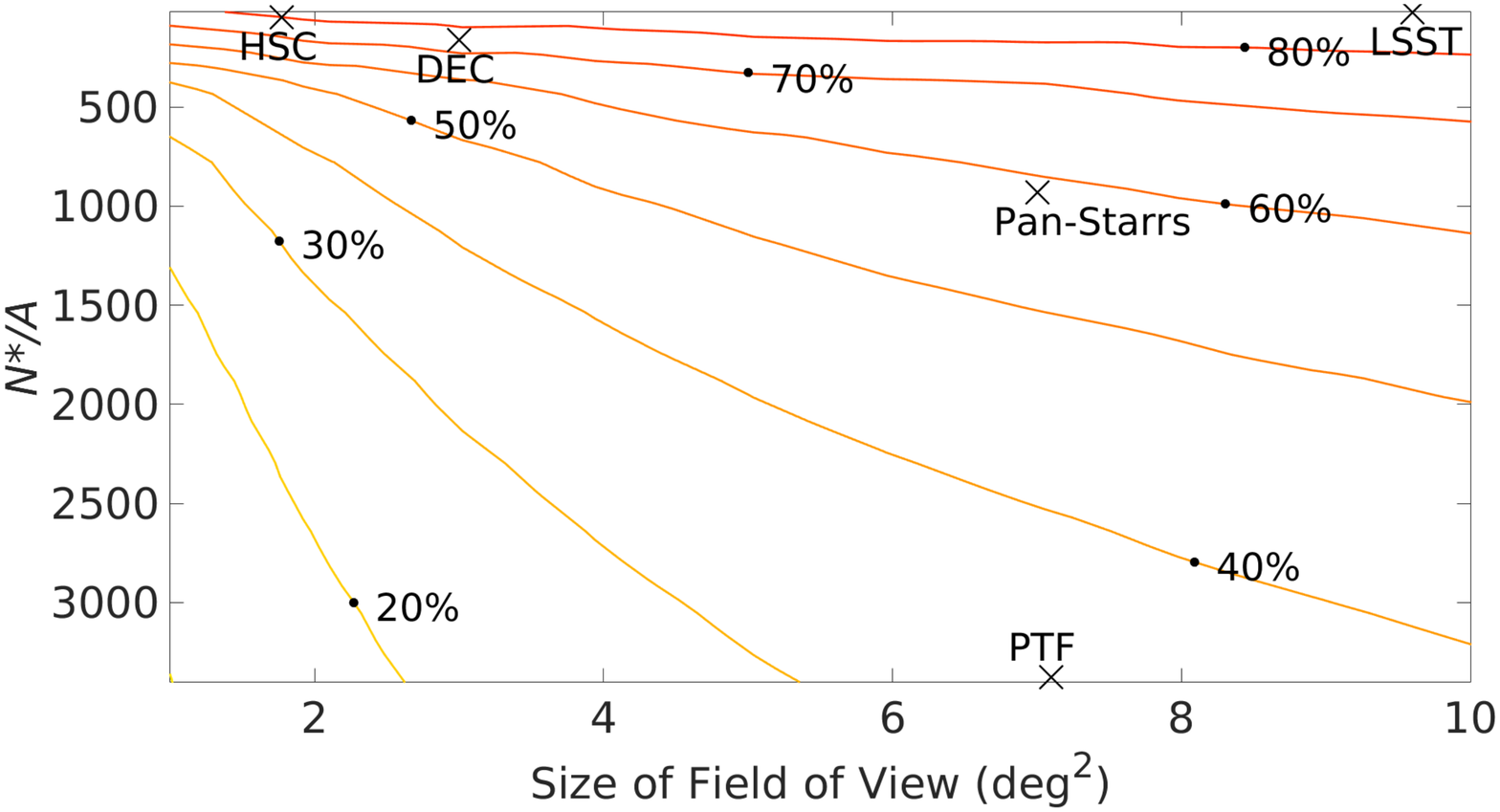}
      \includegraphics[width=.45\textwidth]{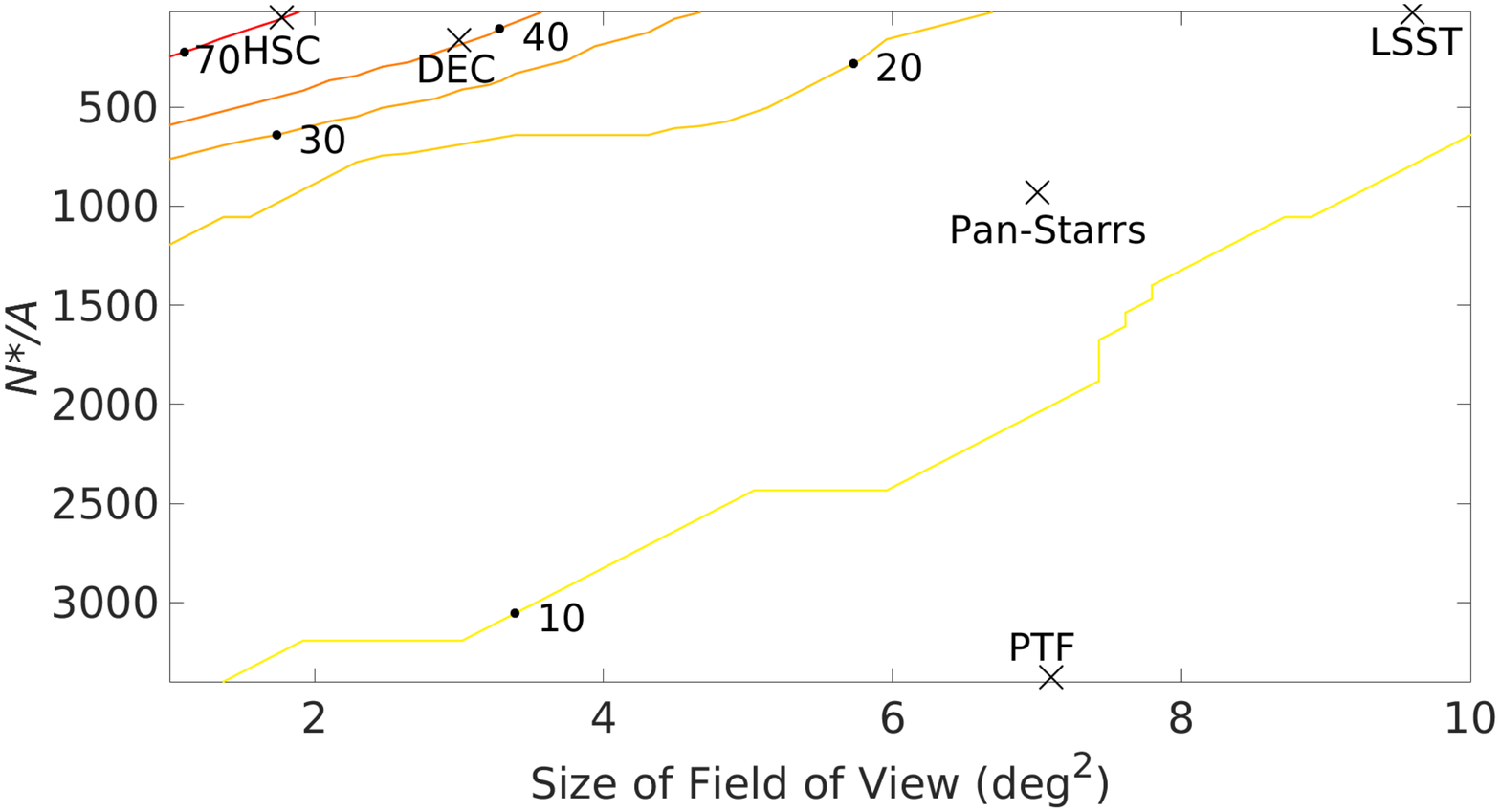}
    }
    \subfigure[The \ac{EM} detection probability and optimized number of fields
for a general telescope observing event 18694 (${\sim}30\,\text{deg}^{2}$).]
    {
      \includegraphics[width=.45\textwidth]{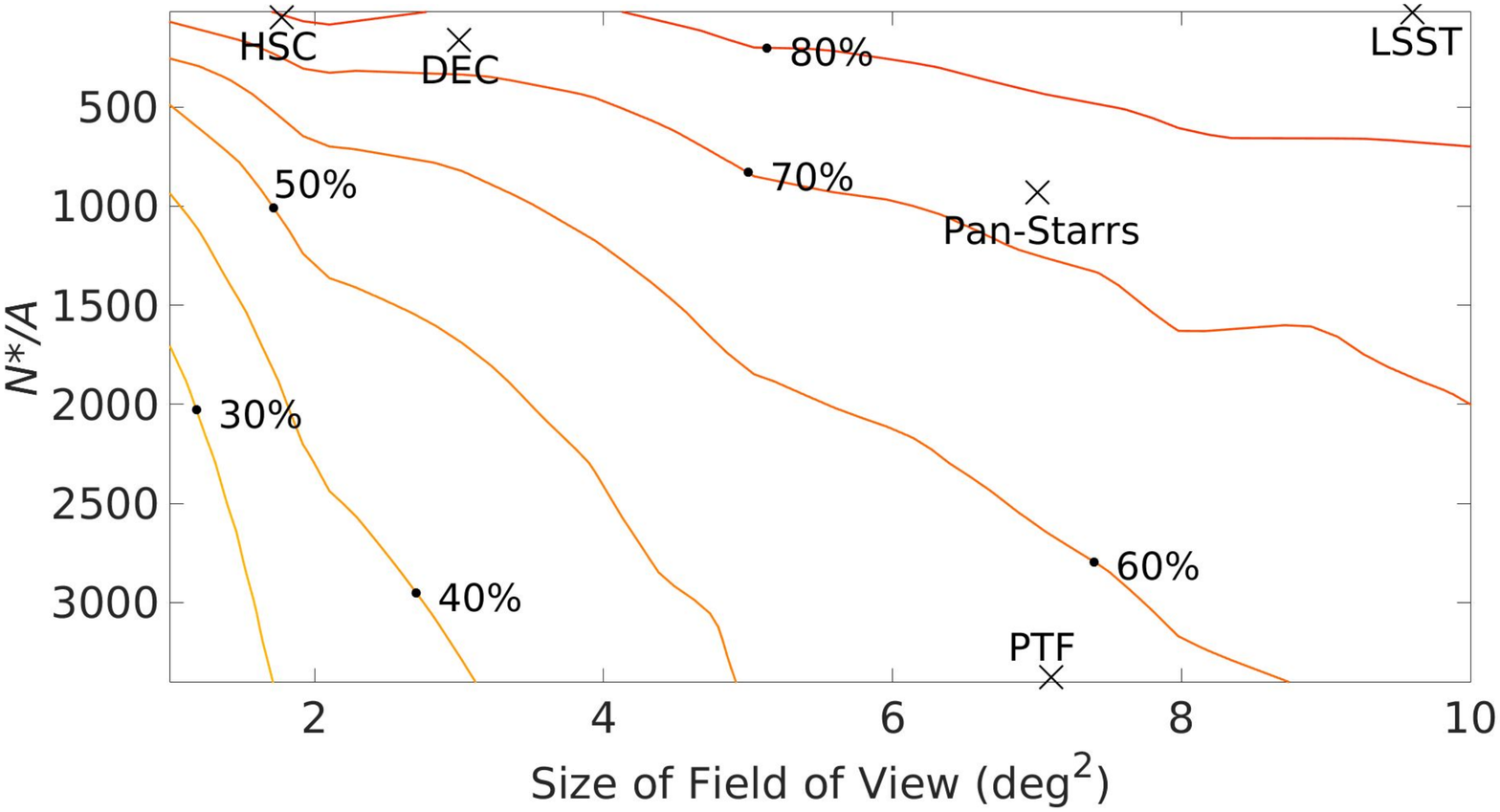}
      \includegraphics[width=.45\textwidth]{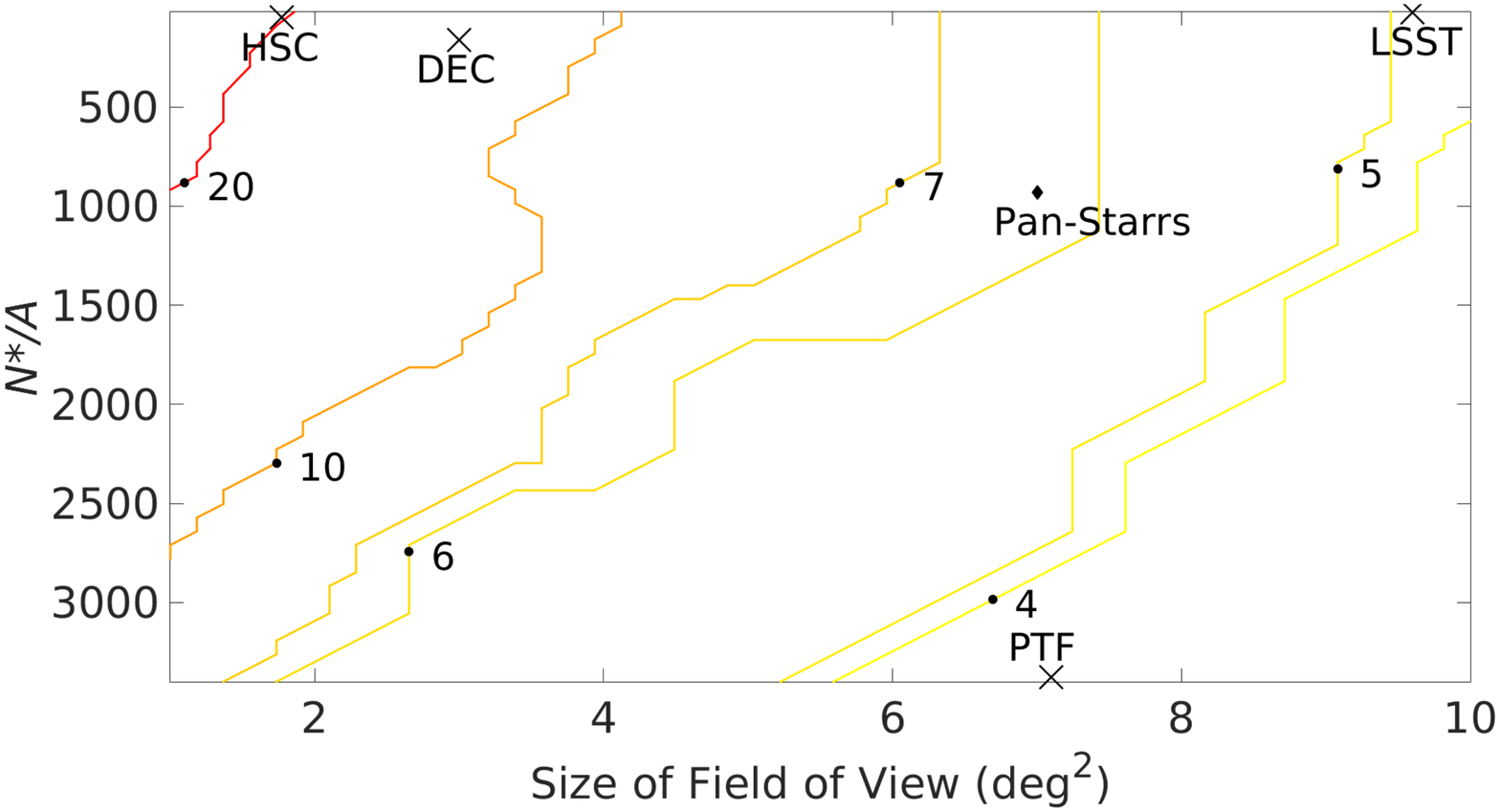}
    }
    \caption{Contours of \ac{EM} follow-up performance of kilonovae as a
function of the size of telescope \ac{FOV} and sensitivity assuming a 6\,hr
total observation. We show results for the 3 simulated \ac{GW} events (a) event
ID 28700, ${\sim}300\,\text{deg}^{2}$, (b) event ID 19296
${\sim}100\,\text{deg}^{2}$, and event 18694 ${\sim}30\,\text{deg}^{2}$. For
each event we plot contours of equal detection probability (left) and the
corresponding optimal number of observing fields (right). Overlayed for
reference on all plots are the locations of the telescopes considered in this
work (including the proposed \ac{LSST}).\label{fig:ref}}
\end{figure*}

%
In Fig.~\ref{fig:ref} we provide an insight into the detection potential of
future telescopes and those not included in our analysis. We have computed the
detection probability $P(D_{EM}|k)$ and corresponding optimal number of fields
$k$ as a function of arbitrary \ac{FOV} and telescope sensitivity. We define
this sensitivity via the quantity $N^{*}/A$, the number of photons per m$^{2}$
required for detection (see Eq.~\ref{eq:td}). For this general case we consider
only a 6\,hr total observation of each of the 3 simulated events.  For
reference we include the 4 telescopes already considered plus the proposed
\ac{LSST}~\citep{abell2009lsst} plotted with points indicating their locations
in the \ac{FOV}, telescope sensitivity plane. The relevant parameters for all
telescopes included are given in Table~\ref{table:tsen}.

%

%
\section{Discussion}\label{sec:discussion}
%
%
The behavior of the detection probability as a function of the number of
observed fields (as shown in Figs.~\ref{fig:re28700},~\ref{fig:re19296},
and~\ref{fig:re18694}) shows that the 2 time allocation strategies produce
similar detection probabilities. In all cases the optimized approach gives
marginally greater probability. As listed in Table~\ref{table:results} we see
that for the particular cases examined in this work the fully optimal approach
leads to a typical gain of a few percent in detection probability over the
equal time strategy. The biggest gains of $\sim 5\%$ are obtained for the
\ac{DEC} telescope. These relatively modest gains suggests that spreading
observation time equally is close to optimal at the optimal number of observing
fields $k^{*}$. We also note that the number of fields at which the peak
detection probability is achieved is similar for both strategies but always
marginally lower for the equal time approach.

%
Both strategies clearly indicate that with a given telescope, there exits a
number of fields at which the probability of a successful \ac{EM} follow-up is
maximized for a given \ac{GW} event and a fixed amount of total observation
time.  In other words, exploring more or fewer fields than necessary will
result in a decrease in detection probability. Although one would expect the
probability of a successful \ac{EM} follow-up first increases with the number
of fields observed, a drop occurs if so many fields are observed that only
observations of short exposure time in some or all of the observed fields are
allowed, given that the observation time is fixed.  This trade-off between
exploring new fields and achieving increased depth within fields is broadly
consistent with ~\citet{Nissanke2012}. Overall, for the same event, the peak
detection probabilities increase as the total observation time increases. For
the same telescope, the detection probability increases as the size of error
region decreases. This is expected since smaller error regions mean that a
telescope requires fewer fields to cover that region at a given confidence. By
examining the curves in Figs.~\ref{fig:re28700},~\ref{fig:re19296}
and~\ref{fig:re18694}, it can be seen that an increase in the total observation
time shifts the position of the peak to larger numbers of fields, but does not
change the general shape of the function.

%
The time allocations shown in the plots on the right of
Figs.~\ref{fig:re28700},~\ref{fig:re19296} and~\ref{fig:re18694} are those
computed for the optimal number of fields. In all cases our optimal strategy
ensures that proportionally more time will be assigned to those fields
containing the greater fraction of \ac{GW} probability (the lower index
fields). In general the range in observation times per field spans $\sim 1$
order of magnitude for a given optimized observation. A surprising feature of
these distributions is that the optimal time allocations can therefore differ
by factors of a few per field with respect to the equal time distribution.
However, both distributions result in very similar detection probabilities.
Recently,~\citet{coughlin2016} have produced an analytic result for the
distribution of observing times under a number of simplifying assumptions
including a uniform prior on peak luminosity. They recommend that the time
spent per field should be proportional to the prior \ac{GW} probability in each
field to the $2/3$ power. We find that our numerical results are broadly
consistent with this result but highlight again that the relative gains in
detection probability are quite insensitive to the time allocation
distribution.  

%
Figure~\ref{fig:ref} shows our results from another perspective, namely the
optimal performance of a any existing or future telescope of arbitrary
sensitivity and \ac{FOV}. As mentioned in Sec.~\ref{sec:maths} our treatment of
the detection threshold criterion $N^{*}$ is simplified and hence these results
should be treated as illustrative rather than definitive. However, as might be
expected, a telescope with poor sensitivity and small \ac{FOV} will be unlikely
to detect an \ac{EM} counterpart unless the \ac{GW} event is particularly well
localized.

%
That \ac{LSST} should explore more fields than \ac{PTF} for these \ac{GW}
events may seem slightly confusing at first glance as \ac{LSST}'s \ac{FOV} is
larger than \ac{PTF}'s. This is because \ac{LSST} is far more sensitive than
\ac{PTF} and can therefore explore as many fields as needed to cover the entire
error region. In comparison \ac{PTF} has to spend a considerable fraction of
its total observation time on each of its observed fields, which results in
\ac{PTF} being only able to observe a limited number of fields.

%
In general, the \ac{EM} detection probabilities achieved are more sensitive to
the telescope parameters for events with smaller error regions. For example,
imagine that a telescope with a $2\,\mathrm{deg}^{2}$ \ac{FOV} and a R-band
limiting magnitude of $20.3$ in $30$~sec ($N^*/A = 2273$) needed to raise the
detection probability $P(D_{\mathrm{EM}}|k)$ by 10\%. For event 18694, where
the size of the $90\%$ credible region is ${\sim}30\,\mathrm{deg}^{2}$, it could
either increase its \ac{FOV} by a factor of $\approx 2$, or reduce its limiting
magnitude to $\approx 21.0$ in $30$\,s. However, if the $90\%$ credible region
is ${\sim}300\mathrm{deg^2}$ as for event 28700, it would have to either further
increase its \ac{FOV} by a factor of $\approx 2.5$ or enhance its limiting
magnitude to $\approx 21.5$ in $30$~sec to have the same factor of improvement.

%
In addition, the fact that at high sensitivity the contours in
Fig.~\ref{fig:ref} appear to be almost flat indicates that as the size of the
\ac{GW} error region becomes larger, the \ac{FOV} of a telescope has negligible
impact on the detection probability $P(D_{EM}|\omega)$. For the design of
future \ac{EM} telescopes performing follow-up observations of \ac{GW} events
there will likely be a trade-off between sensitivity and \ac{FOV}. This result
implies that for \ac{GW} triggers with relatively large error regions,
sensitivity, rather than \ac{FOV}, is the dominant factor determining \ac{EM}
follow-up success. However, we remind the reader that this result is based on
particular choices of source and corresponding prior on the source luminosity
(see Eq.~\ref{eq:magnitude}). Different choices may have different impacts on
the outcome of our method.

%
\section{Future Work}\label{sec:fud}
%
%
This work considers source sky error regions based solely on the information
obtained from low latency \ac{GW} sky localization~\citep{Singer2015:Bayestar}.
One may also include galaxy catalogs to further constrain source locations
within our existing Bayesian approach. In this work, the \ac{GW} source
distance was assumed to be statistically independent of its sky location. In
the future we plan to use more realistic distance
information~\citep{2016arXiv160307333S} therefore enhancing the effectiveness
of our follow-up optimization strategy.  

%
Moreover, since telescopes are distributed at various latitudes and longitudes
on the Earth, different telescopes are able to see different parts of the sky
at different times. As has been studied by~\citep{Rana:2016ve} we would include
the effect of observation prioritization when considering the diurnal cycle and
for instances where \ac{GW} sky error regions may pass below the horizon during
a follow-up observation. Depending on the type of telescope considered we would
also incorporate factors such as the obscuration of the source by the Sun
and/or moon.

%
In this work, we ask the question of detecting kilonovae rather than
identifying and characterizing them. The task of source identification is more
demanding and will require the ability to differentiate our desired sources
from contaminating backgrounds such as SNe and M-dwarf flares.  One way to
accomplish this is to perform multiple observations of the same fields.  In
this case a tentative detection is followed by an observation of the candidate
for a period of time until the source's light curve allows it to be classify as
a kilonova or contaminating noise~\citep{Cowperthwaite2015, Nissanke2012}.  As
mentioned in Sec.~\ref{sec:maths}, simply repeating our proposed observations
would enable the identification of variable objects for deeper follow-ups.
However, a more involved procedure could incorporate light-curve information
into our strategy therefore jointly optimizing the pointings used in both the
detection and identification stages. This more complex strategy is left for
future implementation.

%
\section{Conclusion}\label{sec:ccs}
%
%
In summary, we have demonstrated a proof-of-concept method for quantifying and
maximizing the probability of a successful \ac{EM} follow-up of a candidate
\ac{GW} event. We have applied this method to kilonovae counterparts but we
emphasize that this method is versatile and applicable to any EM counterpart
model. We have shown that there exists an optimal number of fields on which
time should be spent, and that observing more or fewer fields will result in a
decrease in the detection probability. This analysis has been based on the
assumptions of a static telescope with unconstrained pointings, a kilonova
source at constant peak luminosity and the independence of statistical
uncertainty between the distance and the \ac{GW} trigger sky location.

%
Our approach takes as inputs the \ac{GW} sky localization information, and the
selected telescope's characteristics. The method selects the observed field
locations with a greedy algorithm and then uses Lagrange Multipliers to compute
the time allocation for those fields based on maximizing the detection
probability. We have tested the algorithm by optimizing the \ac{EM} follow-up
observations of the \ac{HSC},~\ac{DEC},~Pan-Starrs, and~\ac{PTF} telescopes for
three simulated \ac{GW} events. By comparing the results of our methods with
the results of equally dividing the observation time amongst the observed
fields, we have shown that both strategies return similar results, with our
method producing marginally larger detection probabilities. 

%
In addition, we have provided estimates for the \ac{EM} follow-up performance
of a general telescope of arbitrary sensitivity and \ac{FOV}. These results
indicate that in terms of telescope design, the likelihood of its success in
the follow-up of kilonova signals is approximately independent of the \ac{FOV}
for reasonably sensitive telescopes.

%
To extend this work, it may be helpful to consider including constraints from
galaxy catalogs on source location. Also, the assumptions that the kilonova
luminosity is constant during the observation period, more realistic treatment
of the source distance, and consideration of the dynamics of the telescope with
respect to the source should also be investigated. Finally, the inclusion of
multiple observations of the same fields should be implemented to help
distinguish kilonovae from contaminating sources.

%
\section*{Acknowledgements}
We thank our colleagues Xilong Fan, who provided insight and expertise that
greatly assisted the research. We also thank Keiichi Maeda, Tomoki Morokuma,
Hsin-Yu Chen, and Daniel Holz for assistance and comments that hugely improved
the manuscript. This research is supported by The Scottish Universities Physics
Alliance and Science and Technology Facilities Council. C.~M. is supported by a
Glasgow University Lord Kelvin Adam Smith Fellowship and the Science and
Technology Research Council (STFC) grant No. ST/ L000946/1.


\end{document}